\documentclass[a4paper]{aa}
\usepackage[latin1]{inputenc}
\usepackage[T1]{fontenc}
\usepackage{graphicx}
\usepackage{txfonts}
\usepackage{upgreek}
\usepackage{verbatim}
\usepackage{natbib}
\usepackage{color}
\usepackage{array}

\newcommand{\hii}{H\scriptsize{II}\normalsize}
\newcommand{\micron}{$\upmu$m}
\newcommand{\lsol}{L$_\odot$}
\newcommand{\msol}{M$_\odot$}
\newcommand{\point}{\cdot}
\newcommand{\ms}{m$\point$s$^{-1}$}
\newcommand{\kms}{km$\cdot${s$^{-1}$}}
\newcommand{\kkms}{K$\cdot$km$\cdot${s$^{-1}$}}

\newcommand{\msoly}{\msol$\cdot$yr$^{-1}$}

\newcommand{\cmcube}{cm$^{-3}$}

\newcommand{\mollinej}[2]{($J $={#1}$-${#2})}
\newcommand{\molline}[2]{\,$({#1}$-${#2})$}
\newcommand{\ttp}[1]{$\times 10^{#1}$}

\newcommand{\vlsr}{\varv_\mathrm{lsr}}

\newcommand{\ndhp}{N$_2$H$^+$}

\newcommand{\hms}[3]{$#1^\mathrm{h}#2^\mathrm{m}#3^\mathrm{s}$}
\newcommand{\dms}[3]{$#1\degr#2\arcmin#3\arcsec$}

\newcommand{\itwt}{IRAS~23385$+$6053}
\newcommand{\ieit}{IRAS~18264$-$1152}
\newcommand{\izef}{IRAS~05358$+$3543}
\newcommand{\wfot}{W43-MM1}
\newcommand{\nsfs}{NGC~7538S}
\newcommand{\drts}{DR21(OH)-S}
\newcommand{\drto}{DR21(OH)}
\newcommand{\idccu}{IRAS~18151$-$1208-MM1}
\newcommand{\idccd}{IRAS~18151$-$1208-MM2}
\newcommand{\idzq}{IRAS~18089$-$1732}
\newcommand{\wtic}{W3~IRS5}
\newcommand{\afqq}{AFGL~490}
\newcommand{\wtta}{W33A}
\newcommand{\afto}{AFGL~2136}
\newcommand{\aftf}{AFGL~2591}
\newcommand{\sofo}{S140~IRS1}
\newcommand{\nsfo}{NGC~7538~IRS1}
\newcommand{\nsfn}{NGC~7538~IRS9}
\newcommand{\HDDO}{H$_2^{18}$O}
\newcommand{\CHTOH}{CH$_3$OH}
\newcommand{\SOT}{SO$_2$}
\newcommand{\CHTOCHT}{CH$_3$OCH$_3$}
\newcommand{\CHTCHO}{CH$_3$CHO}
\newcommand{\CTHSCN}{\mbox{C$_3$H$_7$CN}}
\def\old#1{{#1}}

\begin{document}
   \title{Tracing early evolutionary stages of high-mass star formation with
molecular lines }

   \subtitle{}

   \author{M.G.~Marseille\inst{1} \and F.F.S.~van~der~Tak\inst{1,2} \and
F.~Herpin\inst{3,4} \and T.~Jacq\inst{3,4}}


   \institute{
	SRON Netherlands Institute for Space Research, Landleven 12, 9747AD
Groningen, The Netherlands\\
   email: \texttt{[marseille;vdtak]@sron.nl}
   \and
	Kapteyn Astronomical Institute, PO Box 800, 9700 AV, Groningen, The
Netherlands
   \and
   Universit\'e Bordeaux 1, Laboratoire d'Astrophysique de Bordeaux, 2 rue de
l'Observatoire, BP 89, 33271 Floirac CEDEX, France\\
	email: \texttt{[herpin;jacq]@obs.u-bordeaux1.fr}
   \and
   CNRS/INSU, UMR 5804, BP 89 33271 Floirac CEDEX, France
   }

   \date{Received 27 October 2009}

  \abstract
  {Despite its major role in the evolution of the interstellar medium, the
    formation of high-mass stars ($M\geq 10$~M$_\odot$) remains poorly
    understood. Two types of massive star cluster precursors, the so-called
    massive dense cores (MDCs), have been observed, which differ in terms of their
    mid-infrared brightness. The origin of this difference has not yet been
    established and may be the result of evolution, density, geometry
    differences, or a combination of these.}
  {We compare several molecular tracers of physical conditions (hot cores,
    shocks) observed in a sample of \old{mid-IR weakly emitting MDCs}
    with previous results obtained in a sample of exclusively mid-IR bright
    MDCs. We attempt to understand the differences between these two types of
    \old{object}.}
  {We present single-dish observations of HDO, H$_2^{18}$O, SO$_2$, and CH$_3$OH
    lines at $\lambda = 1.3 - 3.5$~mm. We study line profiles and
    estimate abundances of these
    molecules, and use a partial correlation method to search for trends in the
    results.}
  {The detection rates of thermal emission lines are found to be very similar
    for both mid-IR quiet and bright \old{objects. The abundances of H$_2$O,
HDO ($10^{-13}$ to $10^{-9}$ in
    the cold outer envelopes), SO$_2$ and CH$_3$OH differ from source to source but independently of their mid-IR flux.} In contrast, the
methanol class I maser emission, a tracer of outflow shocks, is found to be
strongly anti-correlated with the 12~\micron\ source brightnesses.}
  {The enhancement of the methanol maser emission in mid-IR quiet MDCs may be
indicative of a more embedded nature. Since total masses are similar between the two samples,
we suggest that the matter distribution is spherical around mid-IR quiet sources but
flattened around mid-IR bright ones. 
In contrast, water emission is
    associated with objects containing a hot molecular core, irrespective of
    their mid-IR brightness. These results indicate that the mid-IR brightness
of MDCs is an indicator of their evolutionary stage.}


\authorrunning{M.G.~Marseille et al.}
\titlerunning{Early evolutionary stages of high-mass star formation}

\maketitle
%

\section{Introduction}

High-mass stars play a significant role in shaping their host galaxies, because of their high
UV luminosity and mechanical input (wind shocks and supernov\ae), but their
formation remains poorly understood. The scaling-up of existing models of low-mass star
formation \citep{shu1993} is unhelpful, because radiation pressure would halt
the accretion \citep[e.g.][]{yorke2002} leading to a final stellar mass of
10~\msol\ at most. To allow the formation of more massive stars, the
quasi-static scenario
has been modified by increasing the mass accretion rate and including turbulence
or
\old{energetic} dynamics \citep{mckee2003,henriksen1997}. Other approaches, such as
competitive accretion in massive protostar clusters
\citep{bonnell1997,bonnell2001} or protostellar mergers \citep{bonnell1998}
may also provide an answer to this problem. The physical conditions in high-mass
star
formation sites indeed correspond to a dense and turbulent medium where a monolithic infall onto a
single massive protostar is not expected. As an example, observed molecular
emission line
widths are always dominated by turbulent velocities ($\varv_\mathrm{T} >
0.5$~\kms, \textit{e.g.}, \citealt{sridharan2002}), and are observed
from large to small scales (see C$^{34}$S lines observed by
\citealt{leurini2007} with interferometry in \object{IRAS 05358+3543}), and are greater than the
speed of sound ($a_s \simeq 0.3$~\kms). In addition, Jeans masses of massive
star formation regions are very low compared to their own masses ($2.5$~\msol\
for a mean temperature of $20$~K, compared to few hundred solar masses,
\textit{e.g.} the Cygnus-X sample by \citealt{motte2007}), implying that a high level of fragmentation may play an important role. \old{On the other hand}, the magnetic field could limit this fragmentation
\citep{wardthompson2007,girart2009}. A scenario of clustered massive
star formation is now favoured, but its major steps have yet to be defined
with new observations and studies. 

Searches for early stages of massive star formation have
revealed a class of objects, called high-mass protostellar objects (hereafter
HMPOs) in the sample of \cite{molinari1996} and \cite{sridharan2002}, which
are relatively extended ($\sim$1\,pc), and contain one or
several clumps called massive dense cores (hereafter
MDCs), as seen in the Cygnus-X sample by \cite{motte2007}. The main
characteristics of MDCs are \old{weak} radio emission from ionized gas \old{(less than a few 10~mJy at 2.5~cm)} despite a
high total luminosity ($L > 10^3$~\lsol), a high mass ($>$ few hundreds of solar
masses), and a high density ($n > 10^5$~\cmcube) in a typical size of $0.1$~pc
($\sim 20\,000$~AU). They exhibit evidence of accretion \old{activity}: molecular
bipolar outflows \citep{beuther2002b}, water and methanol masers
\citep{beuther2002d}, and large-scale inflows \citep{motte2003,herpin2009}. 
As reviewed by \citet{zinnecker2007}, MDCs
can be seen as proto-clusters that host the formation of high-mass stars.

MDCs are usually divided into mid-infrared
'quiet' and 'bright' sources (hereafter mIRq and mIRb
sources). \cite{vandertak2000a} define the division between them to be 10~Jy
(12~\micron\ flux density), whereas \cite{motte2007} assume a limit of 10~Jy for
the MSX flux densities at
$21.6$~\micron, in accordance with the distance of Cygnus-X. These values
are equivalent to the 
classical limit between class 0 and class I low-mass protostars obtained from the
ratio of mid-IR 
to total luminosity \citep{andre1993}, as MDCs have similar total
luminosities.
The link between mIRq and mIRb sources is not clear and the various
mid-IR flux densities may be the consequence of different source
orientations relative to the line of sight, or differences in the gas
density distribution \citep{vandertak2006,marseille2008}. This last hypothesis
is supported by the observation of developed hot molecular cores (HMCs 
hereafter) in MDCs
\citep[\textit{e.g.}][]{vandertak2000b,motte2003,leurini2007a}. On the \old{other hand},
chemical data indicate that mIRq sources have a lower temperature than bright
ones \citep{marseille2008}, which imply they represent a less evolved stage.

To more clearly define the evolutionary \old{paths} of mIRq and mIRb sources, this paper
measures chemical, physical, and dynamical tracers in a sample of ten
\old{mid-IRq MDCs}, extending the study by
\cite{vandertak2006} of mIRb-MDCs. We include water species (HDO and \HDDO),
which can exhibit an abundance enhancement because of their release 
in the gas phase from the dust grain surfaces, 
by means of a hot core, or shocks induced by bipolar outflows
(\textit{e.g.} \citealt{cernicharo1990,vandishoeck1996,harwit1998,boonman2003}).
Thus we 
probe the processes critically related to the formation of massive
stars. To further investigate this approach, we focus on \old{to} hot cores and
shock tracers using observations of a high-energy transition of \SOT\ and a
methanol class I maser. We apply \old{statistical} tools to our sample
(correlation
and partial correlation factors) to constrain biases and real trends in our
results.

The paper is arranged as follows. Section~2 presents the source sample, and
Sect.~3 the new observations of HDO, \HDDO, \SOT\ and \CHTOH\ molecular
\old{transitions}.
Section 4 checks for detections and describes the line profiles. 
Section 5 treats the case of the methanol class I maser emission, extracting the
\old{maser}
emission \old{from} the thermal emission. 
In Sect.~6, a modelling method is applied to extract
molecular abundances, themselves analysed  in Sect.~7 using
statistical methods to
find biases and real correlations between them and the physical characteristics of
MDCs. 
In Sect.~8, we discuss our results and present conclusions about the class~I 
maser emission behaviour, describe the differences and similarities between
mIRq- and mIRb-MDCs, and present new ideas about their evolutionary stages.

\section{Source sample}

We observed ten MDCs \old{with a low mid-IR brightness} (see
Table~\ref{tab:source}; fluxes have been corrected to assume a single distance of 1.7~kpc), taken from well-known samples
\citep[][]{beuther2002a,molinari1996,motte2007}. Their physical properties incorporate a large range of masses ($200$-$2000$~\msol) and luminosities (between
$2$\ttp{3} and $3$\ttp{4}~\lsol) within a typical size of $\sim 0.1$~pc. All are
dense
($n_\mathrm{H_2} \simeq 10^5 - 10^7$~\cmcube), \old{most of them ($8$ in total) exhibiting
water and/or methanol class II  masers, and bipolar outflows}. Chosen MDCs \old{are weak at cm wavelengths} (\old{except} for \izef, which exhibits an UC-\hii\ region), making them good candidates to represent an early stage of
the formation of massive stars. Other studies infer a high accretion rate for
these sources ($\sim 10^{-3}-10^{-4}$~\msoly) derived from the power of their
molecular outflows, and a clustered formation process observed by
interferometry \citep{zapata2005,leurini2007,molinari2008,bontemps2009}. 
According to IRAS or MSX data, these sources emit between $0.2$
and $22$~Jy at
12~\micron. This sample is complementary to the one observed by
\cite{vandertak2006}, which contained mIRb-MDCs exclusively.

\begin{table*}[t!]
\begin{minipage}[t]{500pt}
\caption{Source sample.}           
\label{tab:source}      
\renewcommand{\footnoterule}{}
\begin{center}
{\scriptsize
\begin{tabular}{clccccccccccc}
\hline
 \# & Source 	& $\alpha$ 	& $\delta$	& $\vlsr$	
& D	& $L$ & $M$ & $r$ & \CHTOH/H$_2$O &$L_\mathrm{12\mu m}$ & Ref.\\ 
 &	& (J2000)	& (J2000)	& (\kms)	 & (kpc)
& (\ttp{4}\lsol)  & (\msol) & (pc) & masers & (Jy) & \\
\hline
\hline
1 & IRAS 05358$+$3543	& \hms{05}{39}{10.4}	& \dms{+35}{45}{19}	& $-17.6$	&  $1.8$  & $0.7$ & $400$  & $0.10$  & yes/yes &  6.29 & 1,2 \\
2 & IRAS 18089$-$1732	& \hms{18}{11}{51.5}	& \dms{-17}{31}{29}	& $+33.8$ &  $3.6$  & $3.2$ & $1000$ & $0.14$  & yes/yes &  27.27 & 1,3 \\
3 & IRAS 18151$-$1208 MM1	& \hms{18}{17}{58.0}	& \dms{-12}{07}{27}	& $+33.4$	&  $3.0$  & $1.4$ & $330$  & $0.13$  & yes/no & 67.89 & 4 \\
4 & IRAS 18151$-$1208 MM2	& \hms{18}{17}{50.4}	& \dms{-12}{07}{55} & $+29.7$	&  $3.0$  & $0.3$ & $230$  & $0.11$  & no/yes & 0.65 & 4 \\
5 & IRAS 18264$-$1152	& \hms{18}{29}{14.3}	& \dms{-11}{50}{23}	& $+43.6$ &  $3.5$  & $1.4$ & $1200$ & $0.13$  & yes/yes & 12.25 & 1,2,5 \\
6 & W43-MM1			& \hms{18}{47}{47.0}	& \dms{-01}{54}{28} & $+98.8$	&  $5.5$  & $2.3$ & $2000$ & $0.12$  & yes/yes & 11.51 & 2,6 \\
7 & \drto			& \hms{20}{39}{00.8}	& \dms{+42}{22}{48} & $-4.5$	&  $1.7$  & $0.2$ & $450$  & $0.13$  & ... & 3.10 & 7 \\
8 & \drts			& \hms{20}{39}{01.5}	& \dms{+42}{22}{04} & $-3.1$	&  $1.7$  & $0.4$ & $200$  & $0.13$  & ... & 0.88 & 7 \\
9 & \nsfs			& \hms{23}{13}{44.5}	& \dms{+61}{26}{50} & $-57.0$	&  $2.8$  & $1.3$ & $400$  & $0.14$  & yes/yes & 4.61 & 8,9 \\
10 & IRAS 23385$+$6053	& \hms{23}{40}{53.3}	& \dms{+61}{10}{19}	& $-49.0$	&  $4.9$  & $1.6$ & $370$  & $0.05$  & .../yes & 41.96 & 1,6,10 \\
\hline
 & & & & & &  \\
\end{tabular}\\
}
\end{center}
\tablebib{
(1)~\citet{beuther2002c}; (2)~\citet{herpin2009}; (3)~\citet{leurini2007a}; (4)~\citet{marseille2008}; (5)~\citet{szynczak2000,edris2007}; (6)~\citet{molinari1996}; (7)~\citet{motte2007}; (8)~\citet{sandell2003};   (9)~\citet{barvainis1989,cernicharo1990,haschick1989}; (10)~\citet{molinari1998II}.}
\tablefoot{
Columns 2--10 present coordinates, velocities in the local standard of rest, heliocentric distances, luminosities, mass estimates and source sizes at millimetre wavelength of the sources, extracted from references given in the last column.
}
\end{minipage}
\end{table*}


\section{Observations}

\begin{table*}[t!]
\begin{minipage}[t]{500pt}
\caption{Molecular lines observed.}           
\label{tab:obs}      
\renewcommand{\footnoterule}{}
\begin{center}
\begin{tabular}{lcccccccc}
\hline
Species 	& Transition	& Frequency	& $E_\mathrm{up}$ & HPBW
& $\eta_\mathrm{mb}$ & Receiver	& $\delta \varv$\tablefootmark{a} &
$T_\mathrm{sys}$\\ 
		& $J_{K_p,K_o}$	 & (GHz) 	& (K)	& (\arcsec)	
& 	         &                          & (m$\point$s$^{-1}$)   &  (K)      
      \\
\hline
\hline
HDO				& $1_{1,0} \to 1_{0,1}$ & $80.5783$ & $47$ &
$31$ & $0.79$ & A100 & $74$  & $140$-$190$ \\
				& $3_{1,2} \to 2_{2,1}$ & $225.8967$ & $168$ &
$11$ & $0.54$ & B230 & $53$  & $750$-$1600$  \\
H$_{2}^{18}$O			& $3_{1,3} \to 2_{2,0}$ & $203.4075$ & $204$ &
$12$ & $0.57$ & A230 & $59$  & $600$-$1300$ \\
SO$_2$				& $12_{0,12} \to 11_{1,11}$ & $203.3916$ & $70$
& $12$ & $0.57$ & A230 & $59$  & $600$-$1300$ \\
CH$_{3}$OH			& $5_{-1,5} \to 4_{0,4}$E & $84.5212$ & $40$ &
$29$ & $0.78$ & B100 & $71$   & $140$-$190$ \\
\hline
& & & &  \\
\end{tabular}\\
\end{center}
\tablefoot{
Columns 2--9 indicate energy level transitions, line emission rest frequencies,
half power beam width (HPBW), main beam efficiency $\eta_\mathrm{mb}$, receiver
name, velocity resolution $\delta \varv$, and system temperature
$T_\mathrm{sys}$.\\
\tablefoottext{a}{Values for the VESPA backend.}
}
\end{minipage}
\end{table*}

We performed observations at the Pico Veleta with the IRAM
30m antenna\footnote{IRAM is an international institute for research in
millimetre astronomy,
co-founded by the Centre National de la Recherche Scientifique (France), the Max
Planck 
Gesellschaft (Germany) and the Instituto Geografico Nacional (Spain)}, in June
$2006$ and
February $2009$. Observations were performed while the weather was good
($\tau_0^\mathrm{atm}\sim 0.07 - 0.6$ at $225$~GHz). We achieved system
temperatures of
$150$-$190$~K for lower frequency transitions and 
$600$-$1600$~K for higher ones (see Table~\ref{tab:obs} for
further details).

We used simultaneously the A100, A230, B100, and B230 receivers to cover five
molecular emission lines: HDO\molline{1_{1,0}}{1_{0,1}} and
\molline{3_{1,2}}{2_{2,1}} at $80.6$~GHz and $225.9$~GHz respectively,
CH$_3$OH\molline{5_1}{4_0} at $84.5$~GHz,
H$_{2}^{18}$O\molline{3_{1,3}}{2_{2,0}}, and SO$_2$\molline{12_{0,12}}{11_{1,11}}
at $203.4$~GHz. We tuned the VESPA correlator to obtain spectral
resolutions of $20$ and $40$~kHz, hence velocity resolutions of between
$50$ and $80$~\ms\ (see Table~\ref{tab:obs} for details). We
observed sources in a single point mode, pointing on the conti\-nuum peak
emission at
millimetre wavelengths. We chose the wobbler-switching observation mode as the
sources are reasonably isolated, using a throw of 4\arcmin\ to the west.
Pointing and
focus calibrations were set on Jupiter and Neptune. The pointing
accuracy was of the order of 2\arcsec.

We performed the data reduction with the software CLASS from the GILDAS
suite \citep{guilloteau2000}.  
After subtracting linear or polynomial baselines, we summed spectra according to
the position and frequency. We converted antenna temperature $T_{a}^{*}$
into main beam temperature $T_\mathrm{mb}$ using the efficiency parameter
$\eta_\mathrm{mb}$ provided by IRAM (see Table~\ref{tab:obs}). We measured
the line parameters (central velocity, width, peak intensity, and line flux) by
fitting one or multiple Gaussian profiles.
\section{Results}

\subsection{Spectral line detections}

\begin{figure}[t]
\centering
\resizebox{\hsize}{!}{\includegraphics{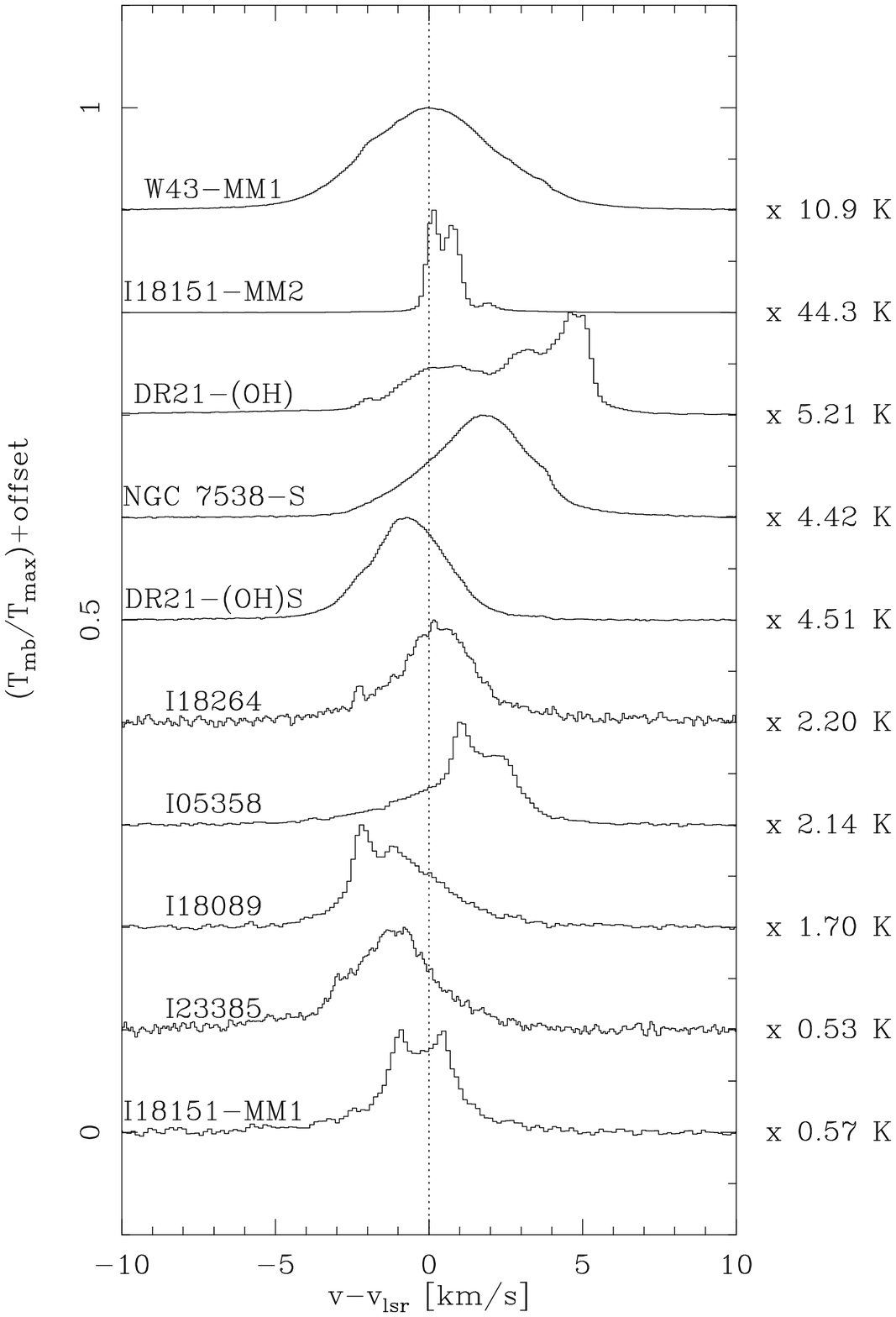}}
\caption{Line profiles of the \CHTOH\ emission observed in our sample. Profiles
have been offset by $\vlsr$ (see Table~\ref{tab:source}) and normalized
to unity with a factor indicated on the right
part of the panel.}
\label{fig:meth-profiles}
\end{figure}

\begin{figure*}
\centering
\begin{tabular}{cc}
\includegraphics[angle=-90,width=7.5cm]{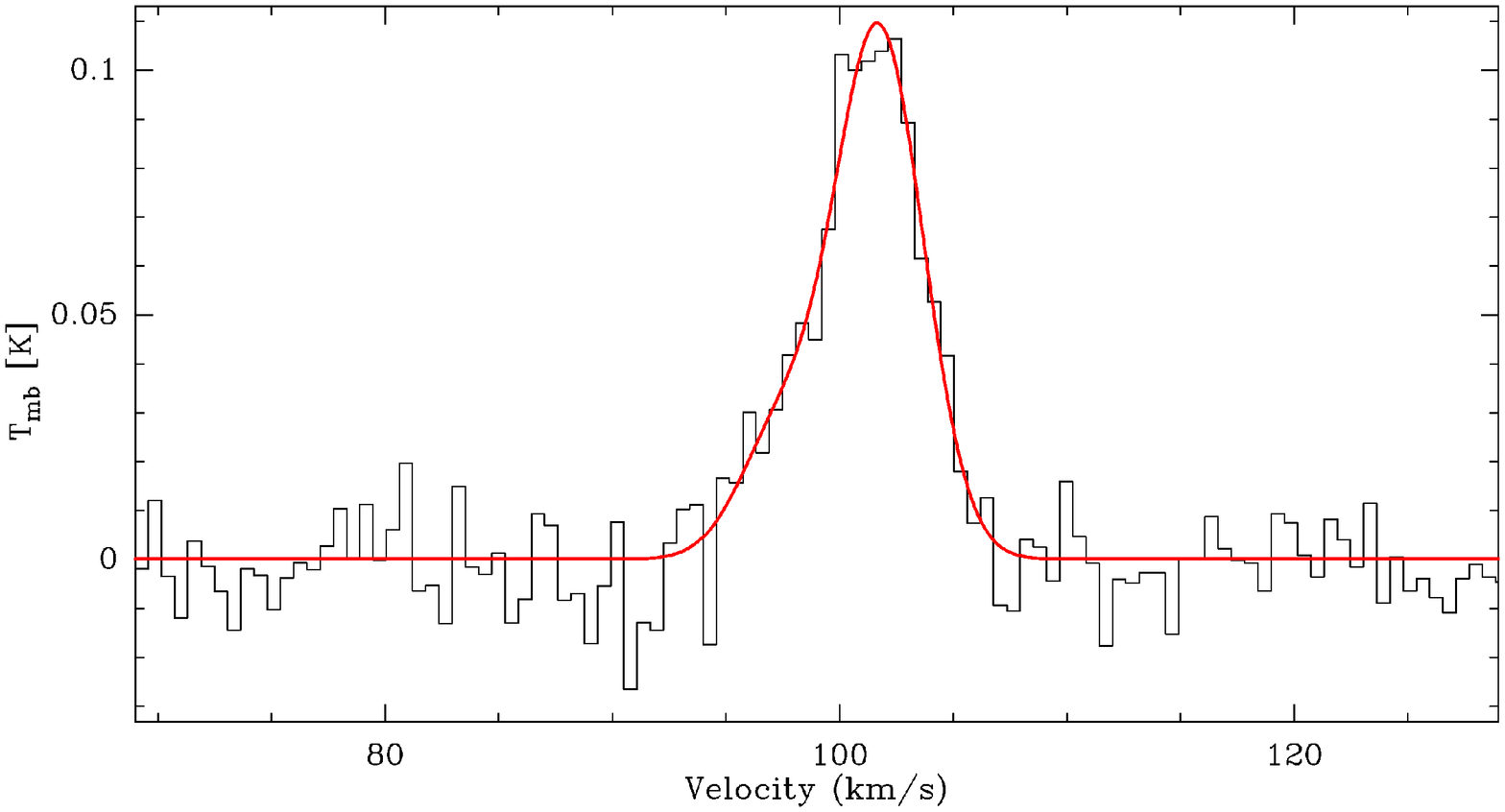} &
\includegraphics[angle=-90,width=7.5cm]{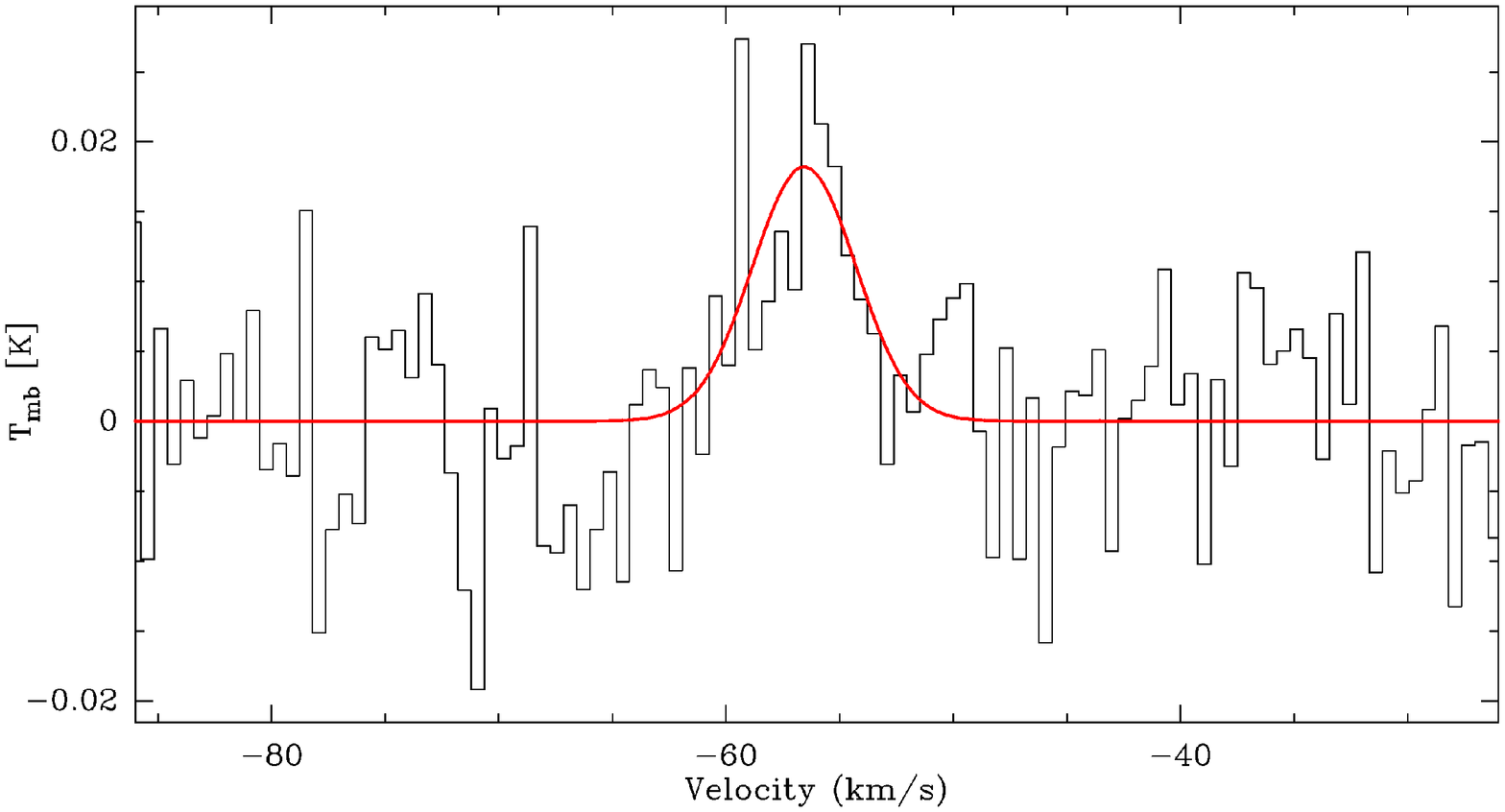} \\
\includegraphics[angle=-90,width=7.5cm]{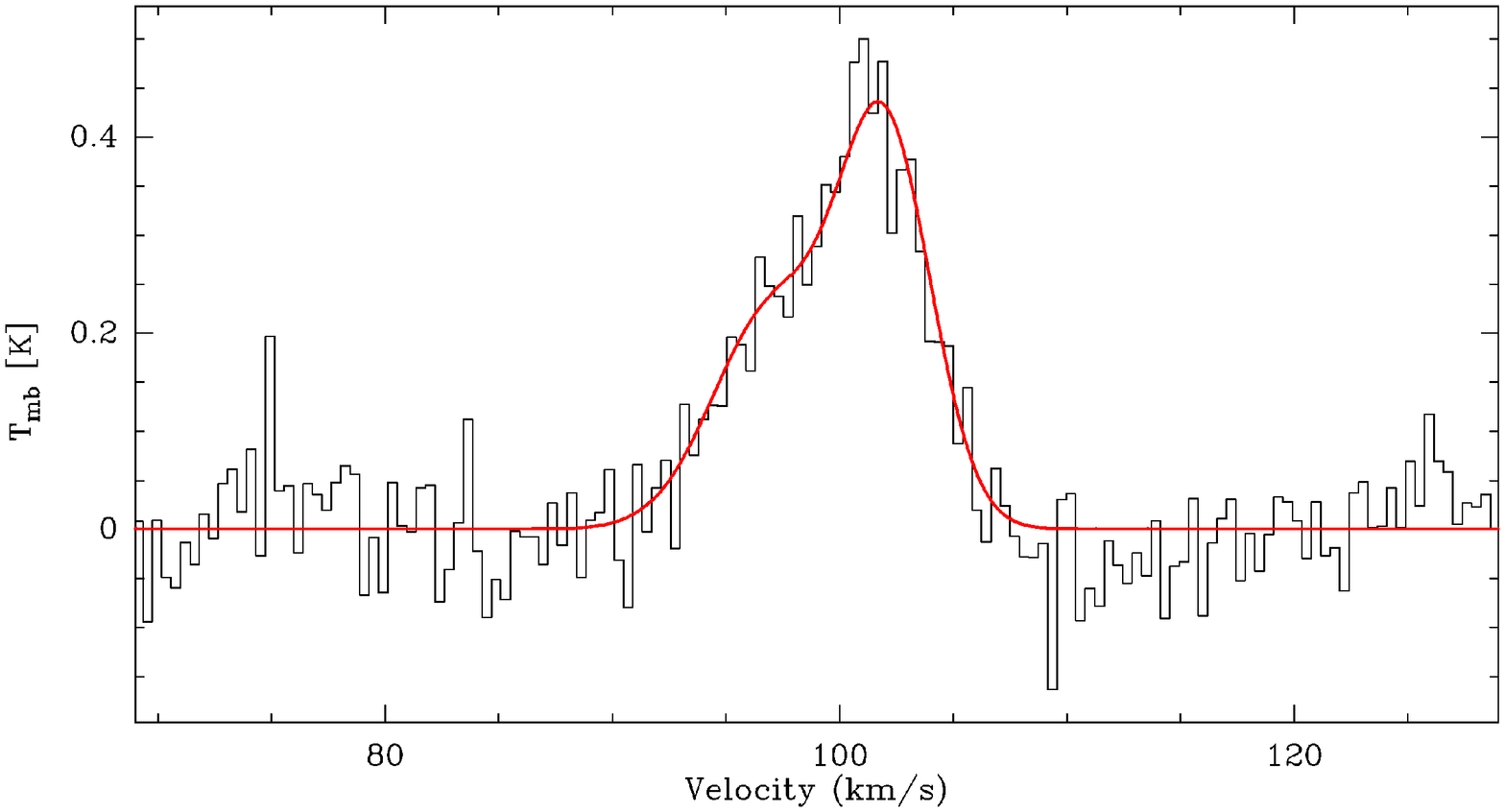} &
\includegraphics[angle=-90,width=7.5cm]{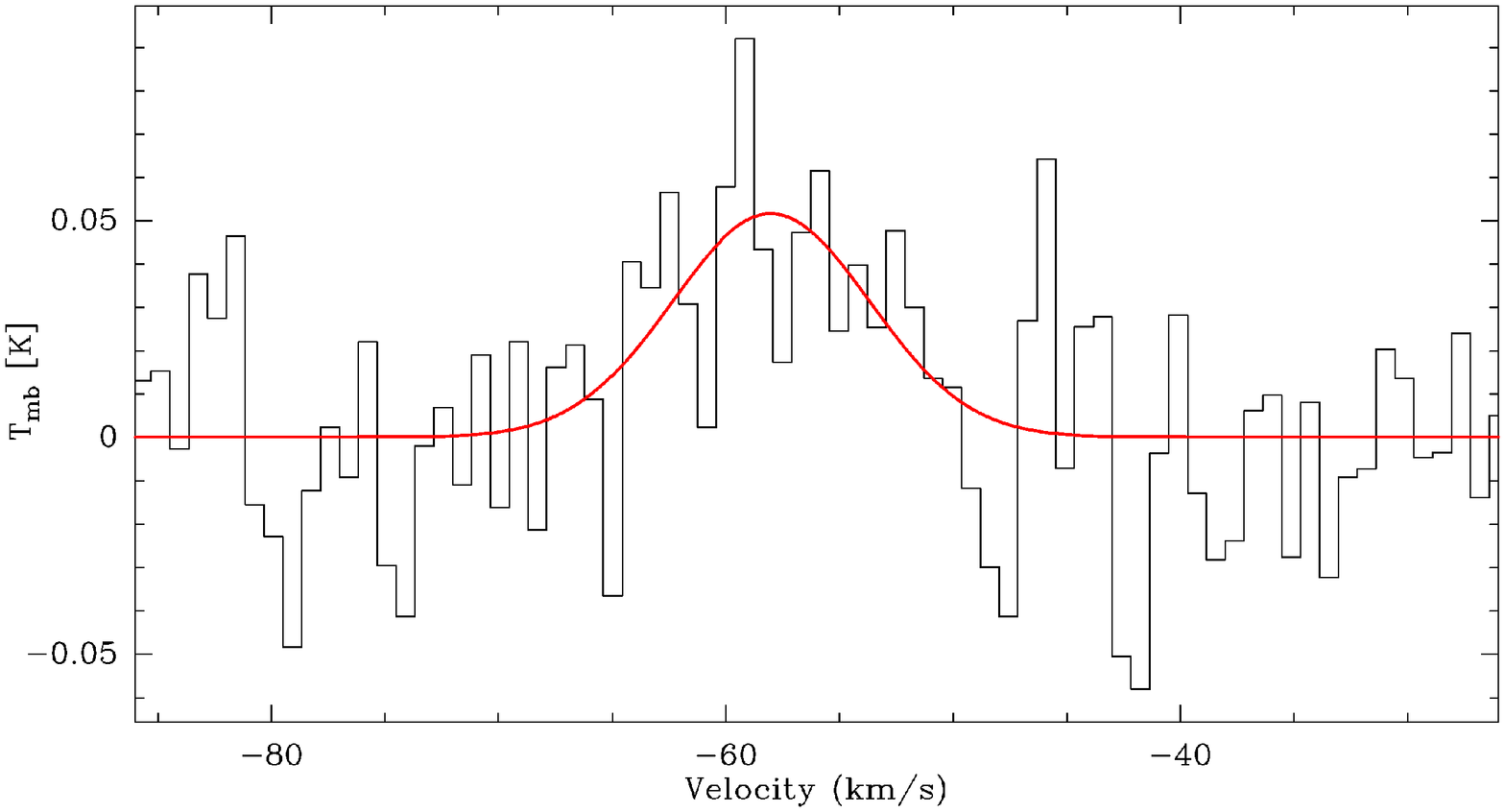}
\end{tabular}
\caption{Observations of the HDO\molline{1}{1} (top) and HDO\molline{3}{2}
(bottom) emission lines from \wfot\ (left) and \nsfs\ (right). For a clearer
view, spectra have been smoothed by reducing the velocity resolution by a factor
2 (see Table~\ref{tab:obs2} for the initial value).}
\label{fig:W$43MM1-HDDO}
\end{figure*}

\begin{figure}[t]
\centering
\resizebox{\hsize}{!}{\includegraphics[angle=-90]{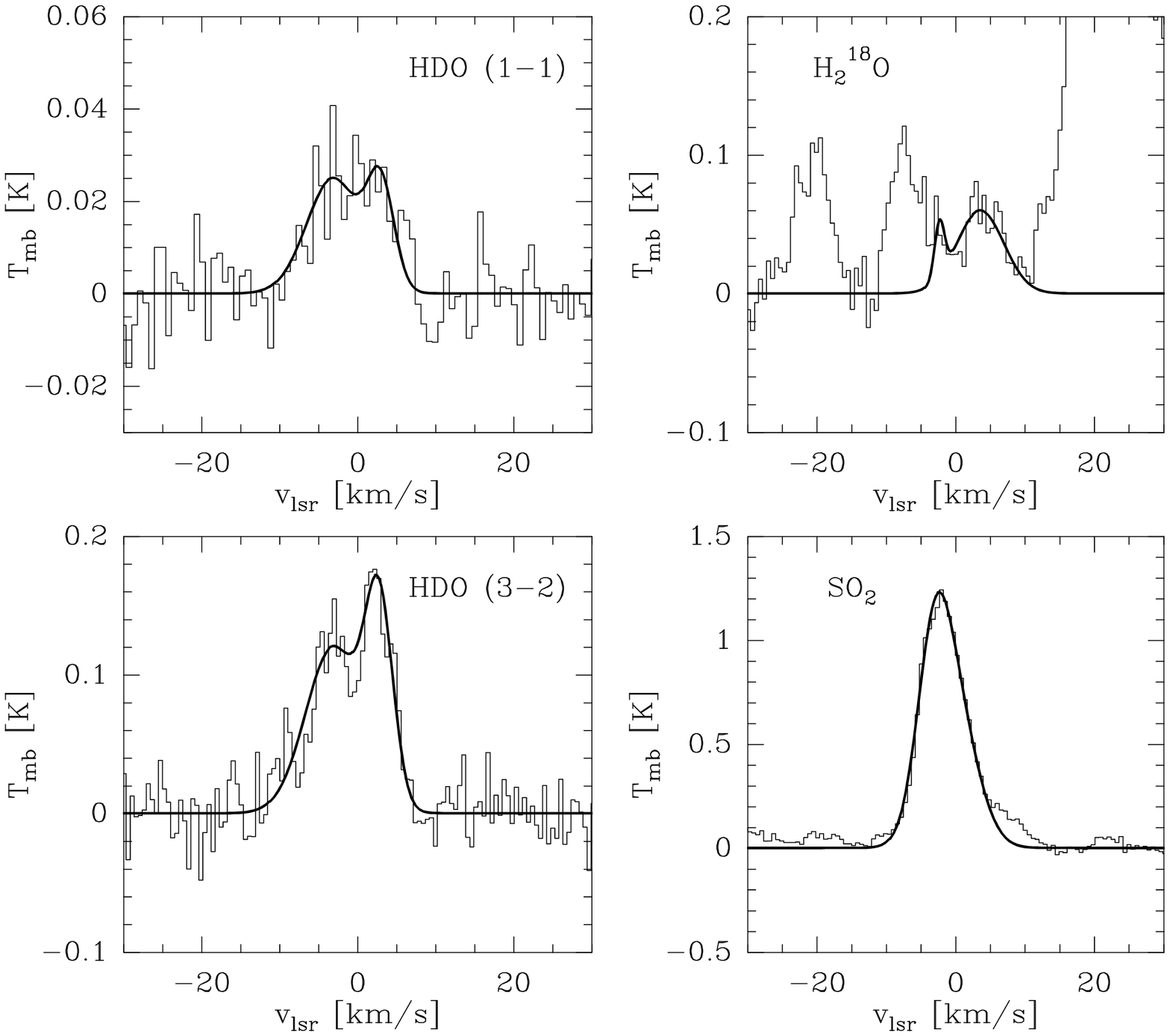}}
\caption{Line profiles of the HDO at 3~mm (top left), 1~mm (bottom left), \HDDO\ (top right),
and \SOT\ (bottom right) observed in \object{DR21(OH)} with their best double-Gaussian fit. Double component fit of \HDDO\ must be considered as tentative to be consistent with HDO line profiles. Profiles are
smoothed by a factor of 5 for a better view.}
\label{fig:drto}
\end{figure}

The methanol emission is detected towards all sources at a
signal-to-noise ratio of 15 or more, with a peak $T_\mathrm{mb}$ varying from
$0.6$ to $\sim 40$~K (see Fig.~\ref{fig:meth-profiles}). For all sources, we
detect wings or multiple velocity
components. Unlike \CHTOH, the \SOT\ lines are detected in 7 sources only and
have a
Gaussian shape. 
The \HDDO\ line is seen toward 3 sources, \wfot, \drto, and \object{IRAS 18089-1732}; the
line profiles exhibit a Gaussian shape, except for \drto\ where the observed signal is more complex
 (see Fig.~\ref{fig:drto} and following section). The HDO line emission at $1$~mm
and $3$~mm is detected
 toward 4 sources: \wfot, \drto, \idzq, and \object{NGC 7538S}. The HDO lines are
always either detected or both non-detected (see Fig.~\ref{fig:W$43MM1-HDDO}). We
do not detect special features in
the line profiles, except for \drto\ whose line profile is most closely fitted by a
two-component velocity model. Measured parameters of single peaked
lines are listed
in Table~\ref{tab:obs2}. Methanol and other special features (such as the
two
velocity components in \drto) are treated in Sect.~\ref{ssec:lin-prof}. 

\begin{table*}[t!]
\begin{center} 
  \caption{Characteristics of observed lines.}
\label{tab:obs2}
{\scriptsize
\begin{tabular}{lllccccc} \hline
Source 	& Species 	& Transition 	& $\varv_\mathrm{lsr}$ 	&
$T_\mathrm{mb}$\tablefootmark{a} & $\Delta\varv$\tablefootmark{b}	& $\int T d\varv$ &
$\sigma_\mathrm{rms}$ \\ 
	&		&		& (\kms)		& (mK)		
& (\kms)	& (\kkms)	& (mK) \\
\hline
\hline
\izef	& HDO           & $J=1_{1,0}-1_{0,1}$           & ...           & ..
& ...	        & ...   & $20$ \\
        & HDO           & $J=3_{2,1}-2_{2,1}$           & ...           & ...
& ... 	      	& ...   & $40$ \\
        & \HDDO\        & $J=3_{1,3}-2_{2,0}$           & ...     	& ...
& ...     	& ...   & $30$ \\
	& \SOT\ 	& $J=12_{0,12}-11_{1,11}$	& $\mathbf{-}15.9(4)$ 
& $90$	
& $6.6(5)$	& $0.61(4)$	& $30$ \\
\idzq	& HDO		& $J=1_{1,0}-1_{0,1}$		& $32.1(1)$	& $130$	
& $4.5(3)$	& $0.62(3)$	& $10$ \\
	& HDO		& $J=3_{2,1}-2_{2,1}$		& $32.7(1)$   	& $420$	
& $6.3(3)$	& $2.8(1)$	& $50$ \\
	& \HDDO\	& $J=3_{1,3}-2_{2,0}$		& $32.7(2)$     & $180$	
& $5.7(2)$	& $1.09(7)$	& $30$ \\
	& \SOT\		& $J=12_{0,12}-11_{1,11}$	& $32.9(4)$     & $430$	
& $9.5(4)$	& $4.4(1)$	& $30$ \\
\idccu	& HDO           & $J=1_{1,0}-1_{0,1}$           & ...           & ...
& ...           & ...   & $20$ \\
        & HDO           & $J=3_{2,1}-2_{2,1}$           & ...           & ..
& ...           & ...   & $50$ \\
        & \HDDO\        & $J=3_{1,3}-2_{2,0}$           & ...           & ...
& ...           & ..   & $40$ \\
        & \SOT		& $J=12_{0,12}-11_{1,11}$	& $33.3(3)$  	& $80$	
& $4.2(8)$	& $0.36(5)$	& $40$ \\
\idccd  & HDO           & $J=1_{1,0}-1_{0,1}$           & ...           & ...
& ...           & ...   & $20$ \\
        & HDO           & $J=3_{2,1}-2_{2,1}$           & ...           & ..
& ...           & ...   & $60$ \\
        & \HDDO\        & $J=3_{1,3}-2_{2,0}$           & ...           & ...
& ...           & ..   & $50$ \\
        & \SOT          & $J=12_{0,12}-11_{1,11}$       & ...     	& ...
& ...      	& ...   & $50$ \\
\ieit   & HDO           & $J=1_{1,0}-1_{0,1}$           & ...           & ...
& ...           & ..   & $60$ \\
        & HDO           & $J=3_{2,1}-2_{2,1}$           & ...           & ...
& ...           & ...   & $330$ \\
        & \HDDO\        & $J=3_{1,3}-2_{2,0}$           & ...           & ...
& ..           & ...   & $290$ \\
        & \SOT          & $J=12_{0,12}-11_{1,11}$       & ...           & ...
& ...           & ...   & $290$ \\
\wfot	& HDO		& $J=1_{1,0}-1_{0,1}$		& $101.3(1)$	& $107$	
& $5.7(2)$	& $0.65(2)$	& $20$ \\
	& HDO		& $J=3_{2,1}-2_{2,1}$		& $100.4(1)$   	& $410$	
& $7.5(2)$	& $3.24(9)$	& $60$ \\
	& \HDDO\	& $J=3_{1,3}-2_{2,0}$		& $100.6(5)$    & $220$	
& $3.8(7)$	& $1.1(3)$	& $80$ \\
	& \SOT\		& $J=12_{0,12}-11_{1,11}$	& $100.7(3)$    & $230$	
& $5.2(7)$	& $1.29(6)$	& $80$ \\
\drts	& HDO           & $J=1_{1,0}-1_{0,1}$           & ...           & ...
& ...           & ...   & $20$ \\
        & HDO           & $J=3_{2,1}-2_{2,1}$           & ...           & ..
& ...           & ...   & $30$ \\
        & \HDDO\        & $J=3_{1,3}-2_{2,0}$           & ...           & ...
& ...           & ..   & $40$ \\
	& \SOT		& $J=12_{0,12}-11_{1,11}$	& $-3.5(1)$  	& $240$	
& $3.5(2)$	& $0.87(3)$	& $10$ \\
\nsfs	& HDO		& $J=1_{1,0}-1_{0,1}$		& $-56.7(5)$ 	& $20$	
& $6(1)$	& $0.11(2)$	& $20$ \\
	& HDO		& $J=3_{2,1}-2_{2,1}$		& $-57.9(8)$   	& $20$	
& $10(2)$	& $0.57(9)$	& $50$ \\
        & \HDDO\        & $J=3_{1,3}-2_{2,0}$           & ...           & ..
& ...           & ...   & $50$ \\
	& \SOT\		& $J=12_{0,12}-11_{1,11}$	& $-55.7(1)$  	& $580$	
& $6.1(1)$	& $3.71(5)$	& $50$ \\
\itwt	& HDO           & $J=1_{1,0}-1_{0,1}$           & ...           & ...
& ...           & ...   & $10$ \\
        & HDO           & $J=3_{2,1}-2_{2,1}$           & ...           & ...
& ...           & ..   & $60$ \\
        & \HDDO\        & $J=3_{1,3}-2_{2,0}$           & ..           & ...
& ...           & ...   & $30$ \\
        & \SOT\ 	& $J=12_{0,12}-11_{1,11}$	& $-50.3(6)$ 	& $40$	
& $8(2)$	& $0.35(5)$	& $10$ \\
\hline
\end{tabular}
}
\end{center}
\tablefoot{
Lines are fitted with a single Gaussian profile. Numbers in brackets indicates the error bar associated with the last decimal written. \drto\ is not reported here, having its own dedicated table.\\
\tablefoottext{a}{Conversion factor is $S/T_\mathrm{mb} = 4.95$~Jy/K for IRAM 30m telescope.}
\tablefoottext{b}{The line emission velocity width is measured at the full width half maximum (FWHM).}
}
\end{table*}  

\begin{figure}
\centering
\resizebox{\hsize}{!}{\includegraphics{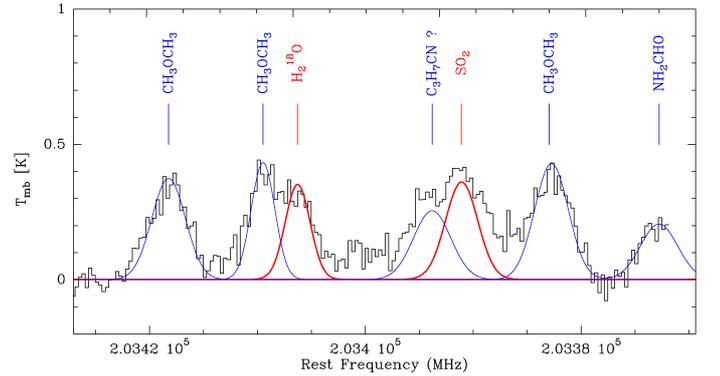}}
\caption{Spectrum of \HDDO\ and \SOT\ emission lines from \wfot, showing the
serendipitous detection of \CHTOCHT, \CTHSCN\ and NH$_2$CHO lines. Spectra are 
smoothed by reducing the velocity resolution by a factor 8 (see
Tab.~\ref{tab:obs2} for the initial value).}
\label{fig:W43MM1-HDDO}
\end{figure}

The 1.3~mm spectrum of \wfot\ shows detections of the expected transitions mixed
up with additional lines that we have identified (see
Fig.~\ref{fig:W43MM1-HDDO}). First, emission line from \HDDO\ is blended with an
emission line of the \CHTOCHT\ molecule at $203.408$~GHz (see
Table~\ref{tab:unex}). We also detect a mixing between the \SOT\ transition and
emission at the frequency of $203.392$~GHz (maybe C$_3$H$_7$CN, as
discovered by \citealt{belloche2009}). This emission seems to be real, as
supported by its velocity width that is similar to
other lines (see Table~\ref{tab:unex}). In the same spectrum, three other
emission lines from \CHTOCHT\ are clearly detected. All the lines have a
Gaussian shape. 
The 3~mm spectrum of \wfot\ shows an unexpected feature near the \CHTOH\ line,
coming 
from the NH$_2$CHO molecule (see Table~\ref{tab:unex}).
One also notes in \idzq\ and \drto\ that the same group of \CHTOCHT\ lines
seen in the \wfot\ spectrum are detected. One of them is blended with
\HDDO.

\begin{table}[t!]
\begin{center} 
  \caption{Characteristics of the unexpected emission lines detected in \wfot.}
\label{tab:unex}
\begin{tabular}{lccccc} \hline
 Species 	& Frequency  	& $E_\mathrm{up}$	&$T_\mathrm{mb}$ 
& $\Delta\varv$\tablefootmark{b}	& $\int T d\varv$  \\ 
		&	(GHz)	& (K)			& (mK)		
& (\kms)	& (\kkms)  \\
\hline
\hline
NH$_2$CHO	& $84.542$	& $7$		& $190$		& $6.51(9)$
& $1.35(2)$ \\
\CHTOCHT	& $203.383$	& $9$		& $430$		& $5.3(5)$
& $2.5(3)$ \\
\CHTOCHT	& $203.406$	& $113$		& $430$		& $3.8(7)$
& $1.8(2)$ \\
\CHTOCHT	& $203.418$	& $9$		& $370$		& $5.3(4)$
& $2.2(3)$ \\
\CHTCHO\tablefootmark{a}	& $211.936$	& $444$		& $260$		& $6(1)$
& $1.6(3)$ \\
\CHTCHO		& $211.957$	& $216$		& $210$		& $6(2)$
& $1.3(4)$ \\
\hline
\end{tabular}
\end{center}
\tablefoot{ 
Numbers in brackets indicate the error bar associated to the last decimal
written.\\
\tablefoottext{a}{Uncertain assignment due to the high level of the upper energy. It could
be C$_3$H$_7$CN, as discovered by \citealt{belloche2009} or a \SOT\
extension of the main emission at $\Delta\varv = -6.2$~\kms\ induced by
outflows.}
\tablefoottext{b}{The line emission
velocity width is measured at the FHWM.}}
\end{table}  

\subsection{Line profiles}

\label{ssec:lin-prof}

For all sources, except \wfot, the methanol
emission line profiles exhibit wings (Fig.~1). Some of them are seen in both the blue and
red-shifted parts (\izef, \object{IRAS 18151-1208 1}, \object{IRAS 18151-1208 2}, \nsfs), and others only on one side
(blue-shifted for \object{IRAS 23385+6053},  red-shifted for \object{IRAS 18264-1152} and \object{DR21(OH)S}). 

In addition to their wings, the methanol lines often show multiple velocity
components. 
This is seen in \izef, \idzq, \idccd, \wfot, and \object{DR21(OH)}. The number of
these components varies between 3 and 4 and their strength from $\sim 1$ to
$40$~K. 

In \drto, the $1.3$~mm HDO transition exhibits a well defined two-component velocity pattern. A double Gaussian fit reproduces the observed line with the parameters reported in Table~\ref{tab:comp-drto}.  The $3$~mm HDO spectrum does not show the two velocity peaks seen at $1.3$~mm. A single velocity Gaussian profile provides a good fit but only with a rather large ($>10$~\kms) line width. We find that a more convincing approach is to analyse this line with a two-line Gaussian model using fixed velocities and widths derived from the 1.3~mm line, obtaining a good result. The fit parameters are reported in Table~\ref{tab:comp-drto}. In the \HDDO\ spectrum, dimethyl ether and \SOT\ lines are clearly detected. A close examination of all dimethyl ether lines reveals that not all lines are well fitted with a single velocity component model because lines are too wide compared to $1.3$~mm HDO components. We choose to fit those lines with a two-component model derived from the peak velocities observed in the methanol spectrum. 
We can then remove from the spectrum the dimethyl ether contribution.
The resulting spectrum shows a signal well above the noise in the range expected for the \HDDO\ transition. The higher velocity part is around the $2.9$~\kms\ velocity observed in the $1.3$~mm HDO spectrum and is well above the noise. It is also well separated from the dimethyl ether line (no blending). On the \old{other hand}, the lower velocity part, around the $-3.2$~\kms component of the HDO line is partly blended by dimethyl ether and the signal-to-noise ratio is \old{lower}. \old{Fit results} are given in Table~\ref{tab:comp-drto}.

To estimate the \HDDO\ emission line strength, and be \old{consistent with the observed HDO line profiles},
the \CHTOCHT\ lines were first subtracted from the spectrum using a model with
two velocity components. The \CHTOCHT\ and water emissions are expected 
to originate in the same region because \old{these} two species are chemically linked.
We then tentatively applied the velocity component values to fit the \HDDO\ emission.
 The resulting line emission parameters are reported in Table~\ref{tab:comp-drto}, showing a blue-shifted
 component that remains doubtful, \old{even if} it cannot be completely rejected (see Fig.~\ref{fig:drto}). The \SOT\
line is detected but a single velocity component fit does not provide a satisfying
profile, where a two-velocity fit constrained by the \old{velocity}
difference (taken from \old{the} methanol velocities) gives a \old{slightly} better result 
(see Table~\ref{tab:comp-drto}). The remaining red wing on the \SOT\ line is mainly related to the dimethyl ether line. The remaining
blue signal on the foot of \SOT\ cannot be identified, but is weak ($70$~mK).

\begin{table}[t!]
\begin{center} 
\caption{Components in the line profile of \drto.}
\label{tab:comp-drto}
{\scriptsize 
\begin{tabular}{lccccc} \hline
Species & Transition & Component &  $\varv$ & $T_\mathrm{mb}$ 	&
$\Delta\varv$\tablefootmark{a} \\ 
 & 	&	&	 (\kms)		& (mK)			& (\kms)	 \\
\hline
\hline
HDO 	& $J=1_{1,0}-1_{0,1}$ &	$1$	& $-3.2(4)$	& $120$		& $8(1)$ \\
	& $J=1_{1,0}-1_{0,1}$ &	$2$	& $2.9(1)$	& $140$		& $4.2(3)$ \\
\hline
HDO	& $J=3_{2,1}-2_{2,1}$ &	$1$	& $-3.2(1)$	& $25$		& $8(1)$ \\
	& $J=3_{2,1}-2_{2,1}$ &	$2$	& $2.9(4)$	& $22$		& $4.2(3)$ \\
\hline
\HDDO	& $J=3_{1,3}-2_{2,0}$ &	$1$	& $-2.4(2)$	& $40$		& $1.7(6)$ \\
	& $J=3_{1,3}-2_{2,0}$ &	$2$	& $3.5(1)$	& $60$		& $8(1)$ \\
\hline
\SOT	& $J=12_{0,12}-11_{1,11}$ &	$1$	& $-3.3(1)$	& $640$		& $5.6(1)$ \\
	& $J=12_{0,12}-11_{1,11}$ &	$2$	& $-0.7(1)$	& $720$		& $8.0(1)$ \\
\hline
\end{tabular}
}
\end{center}
\tablefoot{
Table presents HDO line components at 3~mm (bottom), 1.3~mm (top), \HDDO, and \SOT\ in \drto. Numbers in brackets indicates the error bar associated with the last decimal written.\\
\tablefoottext{a}{The line emission velocity width is measured at the FWHM. Values given for the \HDDO\ blue-shifted component are extracted for the best fit Gaussian function, which remains quite uncertain.}}
\end{table}  

\section{Methanol maser emission}
\label{sec:methmas}

The \CHTOH\,\mollinej{$5_{-1,5}$}{$4_{0,4}$E} transition is often detected as a
so-called class I maser. Unlike class II masers, which originate in the
immediate protostellar environment ($\lesssim$10$^3$~AU), class I masers arise
at some distance
($\gtrsim$10$^4$~AU) from the protostar. 
Here, shocks between jets and ambient gas
cause local density and temperature enhancements where population inversions may
occur \citep[\textit{e.g.}][]{minier2005}. As a consequence, the maser emission
lines are very sensitive to the local environment in which they appear and typically observed to have sharply peaked profile that can be
distinguished from
thermal, turbulent, or outflow emissions.

In our observations, the line profiles show that the maser and thermal emissions
are mixed in different proportions. As we aim to trace shocks, we need to extract maser emission to ensure a reliably determination of any
relation between the supposed evolutionary stages of our sources and \old{the strengths of
the shocks}
that may occur inside. An identical analysis is performed with profiles obtained by
\cite{vandertak2006} to extend our study. 

\subsection{Extraction method}

To disentangle the thermal and maser contributions to the 
\CHTOH\ line profiles, we use the extraction method of \cite{caswell2000}. The strong turbulence within these sources is also relevant a role too, \old{leading to a larger width of the Gaussian line  profile. This is why we enclose its effect under the term "thermal" used in this paper}.
We fit the MDC's envelope contribution in each
methanol profile by including \old{the sources} previously \old{observed} by \cite{vandertak2006}.
 The thermal emission is then fitted by a Gaussian component centred
on the source velocity, an assumption motivated by all
observations of class II or class I methanol masers (\textit{e.g.}
\citealt{caswell1995}, \citealt{valtts2000}, \citealt{blaszkiewicz2004}) exhibiting maser \old{line profiles} that are \old{sharply} peaked and differ from a Gaussian \old{shape}. 
If the line becomes saturated, the maser profile is of course no longer narrow and the wings
grow exponentially \citep{elitzur1992} but this
will never produce a \old{"thermally-looking"} profile. The possibility of blends of
emission originating in multiple maser spots within the telescope beam is also
real, and our assumption cannot be definitely verified until future
interferometer observations are made.

Once the maser emission has been extracted, we measure its velocity-integrated
area $A_\mathrm{maser}$. We also compute the area $A_\mathrm{therm}$ of the
thermal emission and derive the ratio $X_\mathrm{m} = A_\mathrm{maser}/A_\mathrm{therm}$
of these two quantities, to determine the sources in which the maser emission dominates. 

\subsection{Results}

The results of the \CHTOH\ class I maser emission extraction are given in
Table~\ref{tab:extract-maser}, including results for the profiles observed by
\cite{vandertak2006}.

The maser emission is dominant ($X_\mathrm{m} > 10$) in two sources, \idccd\ and \drto. 
The very high emission intensity in these
objects ($T_\mathrm{peak} = 44.3$ and 5.21~K, respectively), and the \old{structure} of their profiles (see Fig.~\ref{fig:meth-profiles}) are consistent with this finding. A group
of four sources (\izef, \idzq, \nsfs, and \itwt) exhibit distinctive maser emission,
($X_\mathrm{m} \sim $1--2). A majority of objects
exhibit weak maser emission ($X_\mathrm{m} \sim 0.1-1.0$): \idccu, \ieit, \wfot, \drts,
\wtta, \aftf, \sofo, \nsfo, and \nsfn. In the three remaining objects, \wtic, \afto, and
\afqq, the maser emission seems to have disappeared and our measurements
must be treated as upper limits (see Table~\ref{tab:extract-maser}).

\begin{table*}[t!]
  \begin{center}
  \caption{Results of the \CHTOH\ class I maser emission extraction.}
  \label{tab:extract-maser}
\begin{tabular}{clccc@{\extracolsep{3pt}}cccc} \hline
   &        &  \multicolumn{3}{c}{Thermal} & \multicolumn{3}{c}{Maser} & \\
\cline{3-5}
\cline{6-8}
\# & Source & $T_\mathrm{th}$ & $\Delta\varv$ & $A_\mathrm{th}$ 
            & $T_\mathrm{m}$ & $\varv_\mathrm{m}$ & $A_\mathrm{m}$ & $X_\mathrm{m}$  \\
   &        & (K)            & (\kms)        & (\kkms)     
            & (K) & (\kms) &  (\kkms)          \\
\hline
\hline
1  & \izef  &	0.48 &	5.0 &	2.55 &	1.7/1.2 &	-16.5/-15.2 &	3.94 &
1.54 \\
2  & \idzq  &	0.59 &	3.6 &	2.26 &	1.5/0.9 &	31.6/32.6 &	3.20 &
1.42 \\
3  & \idccu &	0.44 &	2.2 &	1.03 &	0.3/0.2/0.1 &	32.4/33.9/31.0& 0.72 &
0.70 \\
4  & \idccd &	0.70 &	5.0 &	3.72 &	43.8/37.4/3.8 &	29.9/30.5/31.6& 639.5 &
172 \\
5  & \ieit  &	1.20 &	3.8 &	4.85 &	1.0/0.4 &	43.8/41.3&	2.68 &
0.55 \\ 
6  & \wfot  &	10.9 &	4.2 &	53.9 &	1.5/1.2 &	102.0/96.8 &	9.53 &
0.18 \\
7  & \drto  &	1.94 &	2.2 &	4.84 &	5.1/3.1/1.5/0.4 & 0.9/-1.2/-3.1/-6.5 &
537.35 & 64 \\
8  & \drts  &	3.50 &	2.2 &	8.19 &	2.4/1.8/0.2&	-4.3/-5.2/-0.8 &
6.78 & 0.83 \\
9  & \nsfs  &	1.81 &	3.1 &	5.94 &	2.8/2.2 &	-55.1/-53.3 &	13.2 &
2.22 \\
10 & \itwt  &	0.25 &	2.6 &	0.79 &	0.4/0.3 &	-50.3/-51.8 &	1.29 &
1.86 \\
\hline
a  & \wtic  &	0.07 &	2.6 &	0.18 &	...\tablefootmark{a} &	... &
<5.4\ttp{-3} &	<0.03 \\
b  & \afqq  &	0.14 &	2.4 &	0.36 &	..\tablefootmark{a} &	... &
<3.6\ttp{-2} &	<0.10 \\
c  & \wtta  &	1.51 &	3.5 &	5.59 &	0.4/0.2 &	35.2/39.5 &	2.74 &	
0.49 \\
d  & \afto  &	0.17 &	2.9 &	0.52 &	...\tablefootmark{a} &	... &
<6\ttp{-4}   &	<0.0012 \\
e  & \aftf  &	0.39 &	2.9 &	1.19 &	0.2/0.1 &	-5.9/-8.6 &	0.46 &	
0.38 \\
f  & \sofo  &	0.43 &	2.1 &	0.95 &	0.2/0.1 &	-5.7/-8.2 &	0.55 &	
0.57 \\
g  & \nsfo  &	0.68 &	2.9 &	1.85 &	0.3 &	-56.2 &	0.51 &			
0.27 \\
h  & \nsfn  &	0.64 &	2.0 &	1.35 &	0.4/0.2	& -56.3/-55.5 &	 1.00 & 0.74 \\
\hline
  \end{tabular}\\
  \end{center}
\tablefoot{
The table presents parameters of the main thermal emission fitted by a Gaussian (temperatures $T_\mathrm{th}$, velocity widths at
FHWM $\Delta\varv$ and area $A_\mathrm{therm}$), the resulting maser emission peaks characteristics (temperatures 
$T_\mathrm{m}$ and velocities $\varv_\mathrm{m}$), and the area ratios ($X_\mathrm{m}$).\\
\tablefoottext{a}{No methanol maser features detected over the $3\sigma_\mathrm{rms}$ noise threshold.}
}
\end{table*}

\begin{figure}
\centering
\resizebox{\hsize}{!}{\includegraphics[angle=-90]{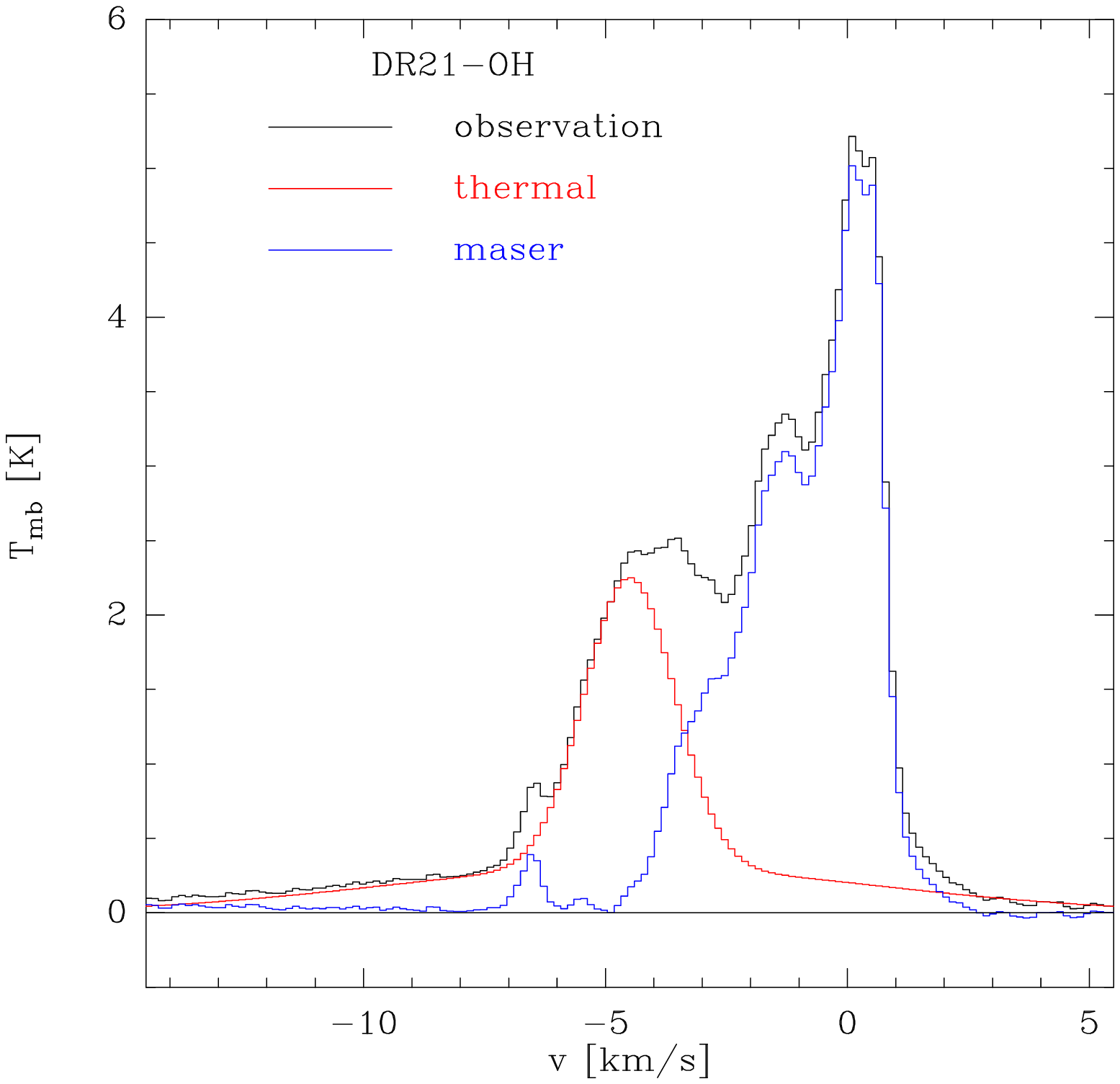}} \\
\resizebox{\hsize}{!}{\includegraphics[angle=-90]{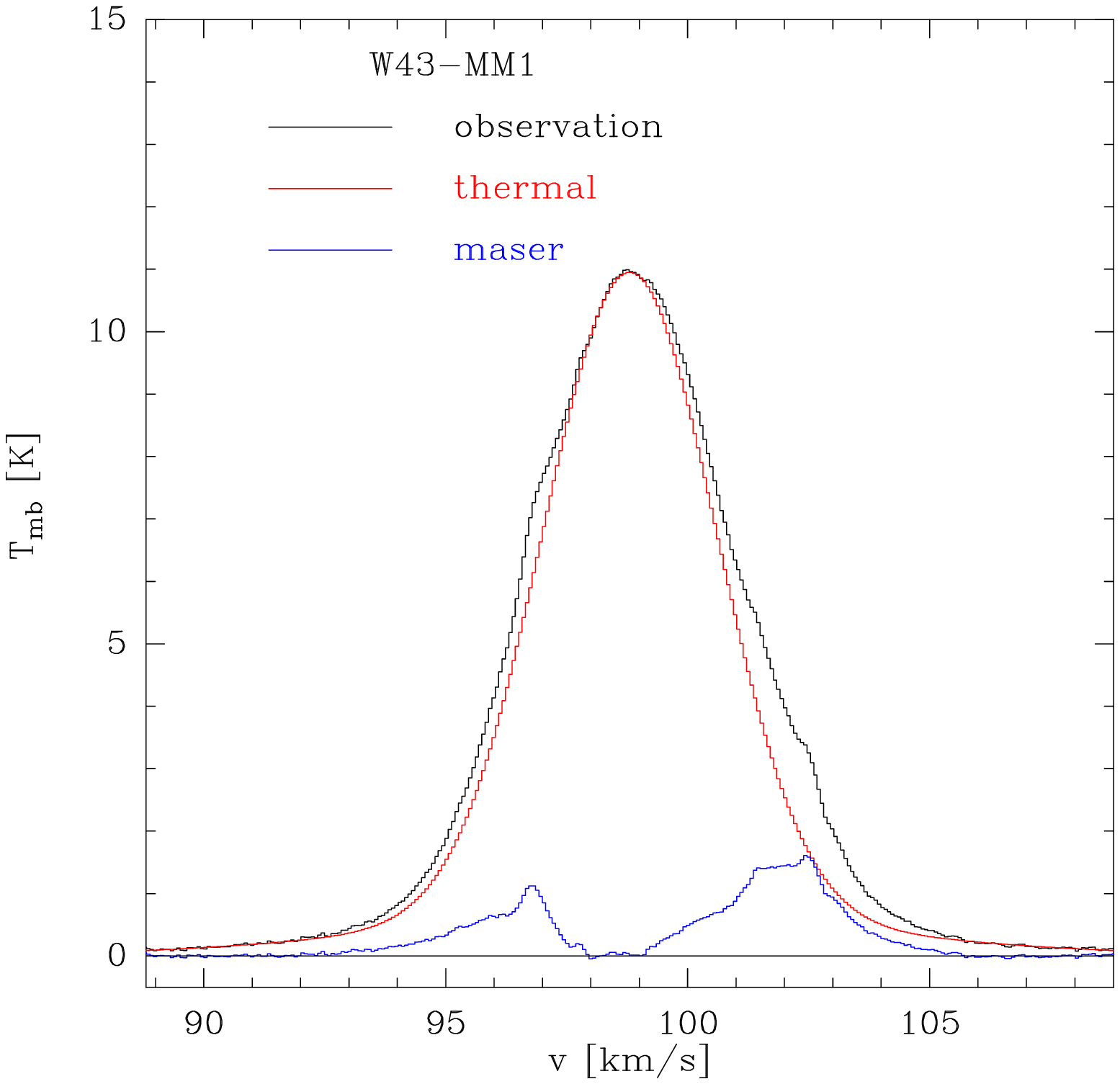}}
\caption{Extraction of the methanol class I maser emission (blue line) from
observations (black line) by fitting the thermal emission with a Gaussian
profile at $\vlsr$ in two sources: \drto\ (top) and \wfot\ (bottom).}
\label{fig:maser}
\end{figure}


\section{Molecular abundances}

\subsection{Modelling method}

To derive molecular abundances from our observations, we use a global modelling
process developed by \cite{vandertak1999} and improved by \cite{marseille2008}.
We first model the SED with the MC3D program \citep{wolf1999} to derive the
mass and the temperature distributions inside the MDCs. The source model,
\textit{i.e.}, its total luminosity,
size, and density distribution is constrained by
observations (see Tables~\ref{tab:source} and \ref{tab:modelpar}), and uses
dust properties derived 
by \cite{draine1984}. We then transfer the density and temperature distribution 
to the RATRAN code \citep{hogerheijde2000}. 
This non-LTE radiative transfer code models the molecular line emissions
observed, and thus we obtain the
molecular abundances and the turbulence level, 
which are the two free parameters of this method.

The density distribution is characterized by a power law of the form 
\begin{equation}
n(r) = n_0 \left(\frac{r}{r_0}\right)^{-p}, 
\end{equation}
where $n_0$ is the density at the reference radius $r_0$ (here 100~AU)
obtained by \old{a} fit \old{to} the optically thin emission of dust at millimetre
wavelengths. The parameter $p$ is derived from continuum maps of MDCs or
is set to be $1.5$ according to a quasi-static infall theory \old{in the inner part of the object} \citep[; see
Table~\ref{tab:modelpar}]{shu1987,beuther2002b}. The
density and temperature distributions are transferred to RATRAN when modelling the
emission lines.
The best fit is obtained for a given gas turbulence velocity
$\varv_\mathrm{T}$ and a molecular abundance relative to H$_2$
($X_\mathrm{mol} = n_\mathrm{mol} / n_\mathrm{H_2}$). When an emission line is
not detected, we derive an upper limit assuming a line velocity width similar to
the \SOT\ one (if available, otherwise we select the thermal component of the
methanol line) and refer to the abundance required for a 2$\sigma_\mathrm{rms}$
signal.

\begin{table*}[t!]
\begin{center} 
  \caption{Modelling parameters.}
\label{tab:modelpar}
\begin{tabular}{clcccccc} \hline
 \# & Source 	& $n_0$  	& $p$	&$\left<n\right>$ 	& $\alpha$
& $\mathbf{T_0}$ & $\left<T\right>$ \\ 
	&	&	(grains.m$^{-3}$)& 	& (cm$^{-3}$)	&	
& (K)	& (K)  \\
\hline
\hline
1 & \izef  & 1.5\ttp{3} & 1.4 & 8.4\ttp{5} & -0.62 & 562 & 31.1 \\
2 & \idzq  & 7.4\ttp{4} & 1.5 & 6.1\ttp{7} & -0.46 & 407 & 36.1 \\
3 & \idccu & 8.7\ttp{2} & 1.2 & 5.5\ttp{5} & -0.60 & 631 & 32.5 \\
4 & \idccd & 7.5\ttp{2} & 1.3 & 6.5\ttp{5} & -0.61 & 427 & 24.2 \\
5 & \ieit  & 4.0\ttp{3} & 1.5 & 9.8\ttp{5} & -0.61 & 676 & 32.6 \\
6 & \wfot  & 2.0\ttp{5} & 2.0 & 5.5\ttp{6} & -0.54 & 302 & 20.0 \\
7 & \drto  & 4.1\ttp{4} & 1.5 & 3.5\ttp{7} & -0.44 & 191 & 20.7 \\
8 & \drts  & 1.8\ttp{4} & 1.5 & 1.5\ttp{7} & -0.43 & 224 & 25.3 \\
9 & \nsfs  & 2.9\ttp{4} & 1.5 & 2.4\ttp{7} & -0.45 & 316 & 30.4 \\
10 & \itwt & 5.4\ttp{5} & 1.5 & 5.0\ttp{8} & -0.42 & 347 & 38.7 \\
\hline
\end{tabular}
\end{center}
\tablefoot{
Table presents the source name, the number of dust grains $n_0$ per meter cube at $r_0 = 100$~AU (following a standard MRN distribution in size and types), the power-law index $p$, the mean $\mathrm{H_2}$ density in the source, the best-fit temperature distribution coefficients ($T = T_0(r/r_0)^\alpha$) and the mean temperature in the source.
}
\end{table*}  

\subsection{Results}

The HDO abundances obtained 
are between $10^{-13}$ and $10^{-9}$ when the line
emission is detected (\idzq, \wfot, \drto, and \nsfs; see
Table~\ref{tab:abundances}). 
For other sources, the upper limits are between
$10^{-13}$ and $10^{-11}$. \itwt\ \old{is an} exception with the high upper limit
of $1.5\times 10^{-9}$ for the HDO\,\molline{3_{1,2}}{2_{2,1}} transition. This
high value (see
Table~\ref{tab:abundances}) is mainly due to the compact shape of the source
($0.05$~pc compared with the size of the other sources, \textit{i.e.} $\sim
0.13$~pc) that increases the beam dilution for \old{high-excitation} lines.
Furthermore, \itwt\ is a very dense and hot source (see
Table~\ref{tab:modelpar}) where a high abundance of water species is needed to
reach the level of the critical density, permitting us to avoid strong
self-absorption of the line emission (particularly for high-energy transitions),
and explaining our high upper limit.

When the emission line of \HDDO\ is detected (\textit{i.e.}
in \idzq, \wfot, and \drto), its abundance is inferred to be
between $10^{-11}$ and $10^{-9}$, giving a
$\mathrm{H_2^{16}O}$ abundance estimation between 5\ttp{-8} and 5\ttp{-7}, 
assuming a solar isotopic ratio $\mathrm{^{16}O/^{18}O} = 500$. Other sources
give an upper limit between $10^{-13}$ and $10^{-12}$, hence a main water
isotope abundance between 5\ttp{-11} and 5\ttp{-10}. This result seems to indicate a
low water abundance in the MDC concerned, relative to the values measured by
\cite{vandertak2006} 
(see Tables~\ref{tab:abundances} and \ref{tab:ratio}), except for \wfot.
The comparison of these abundances with those obtained with the high-energy
transition of HDO 
($E_\mathrm{up} = 168$~K, hence close to the \HDDO\ transition considered) show
that they are
consistent and higher than the low-energy transition of HDO ($E_\mathrm{up} =
47$~K), \old{except \wfot\ } (see Sect.~8 for a discussion of this result). 

The HDO and \HDDO\ abundances that we derive permit us to estimate the
HDO/H$_\mathrm{2}$O ratio in MDCs. Assuming that
$\mathrm{^{16}O/^{18}O} = 500$, the ratios are 33\ttp{-4}, 7\ttp{-4}, and
19\ttp{-4} for \idzq, \wfot, and \drto, respectively. These values are obtained
from high-energy transition results (see Table~\ref{tab:ratio}).
The D/H ratio agrees with the values obtained
by \cite{vandertak2006}, giving values between 3\ttp{-4} and 38\ttp{-4}. Our
results and previous ones are reported in
Table~\ref{tab:ratio}.

The \CHTOH\ abundances are derived from the thermal emission components and are 
spread between $10^{-10}$ and $10^{-8}$, \textit{i.e.} of the
same order of magnitude as those derived by \cite{vandertak2006} for mIRb MDCs.
We note
that \wfot\ has the highest abundance in the sample, with a value reaching
$5.4$\ttp{-7}.

The modelling of the \SOT\ emission line implies an abundance of between $10^{-12}$
and $10^{-10}$ (see Table~\ref{tab:abundances}), which confirms previous
estimations in MDCs \citep{vandertak2003,wakelam2004a}. In particular, \wfot,
\izef, and \ieit\ were already studied in \cite{herpin2009}, where a
multi-transition evaluation of abundances is performed. For these sources, our
results agree with their estimations. Even in \ieit,
where an upper limit has been derived, our result is in accordance with the
abundance estimated in their study. 

We derive turbulent velocities higher than the speed of sound
($\varv_\mathrm{T} = 0.85-2.9$~\kms\ compared to $a_\mathrm{s} =
0.3-0.5$~\kms), confirming this MDC characteristic.

\begin{table*}[t!]
  \begin{center}
  \caption{Modelling results.}
  \label{tab:abundances}
\begin{tabular}{lccccc|ccccc} \hline
	&	\multicolumn{5}{c}{$X_\mathrm{mol}$} &
\multicolumn{5}{c}{2$\varv_\mathrm{T}$ (\kms)} \\
 Source  & \multicolumn{2}{c}{HDO} & \HDDO & \SOT & \CHTOH &
\multicolumn{2}{c}{HDO} &
\HDDO & \SOT & \CHTOH \\
	&	3~mm & 1~mm	   &    &  & & 3~mm & 1~mm	   &      & & 
\\ 
\hline
\hline
\izef  &  $<3.8(-12)$  &  $<7.6(-13)$ & $<4.8(-13)$ & $5.1(-11)$ & $1.3(-9)$ &
... & ... &
... & 2.4 & 2.9 \\  
\idzq  &  $2.0(-11)$  &  $1.3(-10)$ & $7.9(-11)$ & $4.2(-11)$ & $1.3(-10)$ & 2.4
& 4.2 & 5.2
& 5.8 & 2.1\\ 
\idccu &  $<5.8(-13)$  &  $<2.2(-13)$ & $<5.2(-13)$ & $1.2(-10)$ & $1.6(-9)$ &
... & ... &
... & 2.5 & 1.4 \\  
\idccd &  $<2.0(-12)$  &  $<1.1(-12)$ & $<1.8(-12)$ & $<1.2(-10)$ & $1.0(-8)$ &
... & ... &
... & ... & $3.2$ \\  
\ieit  &  $<2.3(-12)$  &  $<1.6(-12)$ & $<1.1(-12)$ & $<1.2(-10)$ & $3.2(-9)$ & ... & ... &
.. & ... & 2.4 \\ 
\wfot  &  $1.5(-9)$  &  $1.7(-9)$ & $4.6(-9)$ & $2.4(-10)$ & $4.8(-7)$ & 2.9 &
2.9 & 2.7 &
3.3 & 4.2\\  
\drto\tablefootmark{a}  &  $6.1(-13)$  &  $8.0(-11)$ & $8.5(-11)$ & $8.4(-11)$ &
$1.0(-10)$ & 2.3 & 3.2 & 1.7
& 3.7 & 1.1\\  
\drts  &  $<8.0(-14)$  &  $<2.0(-11)$ & $<1.8(-13)$ & $7.8(-12)$ & $4.0(-10)$ &
... & ... &
.. & 2.0 & 1.2 \\ 
\nsfs  &  $8.9(-14)$  & $1.3(-13)$ & $<2.8(-13)$ & $3.4(-11)$ & $4.5(-10)$ & 2.8
& 2.6 & ...
& 3.5 & 1.9 \\ 
\itwt  &  $<1.5(-13)$  &  $<1.5(-9)$ & $<1.5(-10)$ & $3.4(-11)$ & $9.4(-10)$ &
... & ... &
... & 4.8 & 1.0 \\ 
\hline
\end{tabular}\\
\tablefoot{
Table presents molecular abundances relative to H$_2$ and turbulent velocities derived from the global modelling method. Values are indicated with the form $x(-y) = x \times 10^{-y}$. When a line emission is not detected, the upper limit
assume a line velocity width similar to the \SOT\ one and refers to the abundance required for a 2$\sigma_\mathrm{rms}$ signal.\\
\tablefoottext{a}{Two velocity components with different line widths.}
}
\end{center}
\end{table*}

\begin{table}[t!]
  \begin{center}
  \caption{Abundance ratios.}
  \label{tab:ratio}
\begin{tabular}{clcc}
\#	& Source	 &  HDO/H$_\mathrm{2}$O & H$_2$O \\
	&		 & \ttp{-4} 		& \ttp{-8} \\
\hline
\hline
2 & \idzq  &  33 	& 4.0\\ 
6 & \wfot  &  7  	& 230\\  
7 & \drto  &  19 	& 4.3\\ 
\hline 
a & \wtic  &  13 	& 30 \\ 
c & \wtta  &  30 	& 80\\ 
e & \aftf  &  3  	& 40\\ 
f & \nsfo  &  38 	& 50\\
\hline
\end{tabular}\\
\tablefoot{
Table presents HDO/H$_\mathrm{2}$O ratios and water abundance derived from
\HDDO. Values in the bottom part are taken  from radiative transfer model
results and \HDDO\ column densities of \citet{vandertak2006}.
}
\end{center}
\end{table}

A large number of sources in our sample are found to have a low upper limit to their \HDDO\
abundances (see Table~\ref{tab:abundances}).  We suspect that most of the water
is frozen in a solid state on the surface on dust grains in these cases. As \old{for} HDO,
\itwt\ \old{is an} exception but this result can be easily explained (see second
paragraph). At the same time, some sources have a clear water abundance
enhancement: results on \wfot\ confirm that it hosts a HMC \citep[already
detected by][]{motte2003,herpin2009}, and others on \idzq\ and \drto\ may
indicate the same.

\section{Correlations}

To investigate how the molecular line emission is related to the physical
properties of the MDCs, we compare the line fluxes and abundances to the mid-IR
(12~\micron) emission seen by MSX and IRAS. Previous studies identified a link
between mid-infrared emission and the physical evolution of MDCs
\citep{vandertak2000a,marseille2008}. To increase the sample size, we include
the data collected by \cite{vandertak2006} for eight mIRb-MDCs.

\subsection{Method}

We compute the correlation factors $\rho_{x,y}$ (see Appendix~\ref{app:corr})
between the following parameters: mass $M$, luminosity $L$, thermal line width
$\delta\varv$ (from the \SOT\ emission or, if not detected, thermal component of
\CHTOH), \old{monochromatic luminosity} $L_{12}$ at 12~\micron, the molecular emission-line
flux observed in \CHTOH\ (masered $A_\mathrm{m}$, thermal $A_\mathrm{t}$, the
ratio $X_\mathrm{m}$, and the abundance X$_\mathrm{mol}$) and the \old{abundance of H$_2$O and HDO} in sources where they are detected. 
Sources with upper limits to the 12\,$\mu$m luminosity or the molecular emission-line flux are not used in the correlation analysis.
Applying a two-tailed correlation test for $N=18$ \old{data points} (or 7
for water), we reject correlation factors with values smaller than $0.49$
(respectively $0.76$) in absolute value. We also use the method of partial
correlations (see Appendix~\ref{app:corr}) to search for biases that could be
introduced by the sample selection \old{and extract the most probable real correlations, this allowing us to correct for false correlations induced by real ones} \citep[see][for a review of the
subject]{rodgers1988}.

\subsection{Search for biases in the sample}

In our systematic search for correlations, we first look for links between
`standard variables' (mass, luminosity, and thermal line width and luminosities at 12~\micron). Apart from the thermal line widths, correlations are
checked in logarithmic scales. To reliably compare our sources, we need to account for their different distances. 
To do so, we assume  that they all are at a reference distance of 1.7 kpc by
applying a correction factor to the flux densities, \textit{i.e.} measuring monochromatic luminosities at 12~\micron\ and comparable methanol maser emissions. No correction is made for beam dilution, as the infrared and maser
emissions are known to originate in a region much smaller than the beam
size (point-like emission).
We search for correlations with `standard variables' regardless of whether there is a physical reason to expect a correlation: this search acts as a reality check on our method.

Three relations are identified when applying the correlation method. First, a link seems to exist
between the mass and the luminosity of the sources
($\rho_{M,L} = 0.50$, thus $97.5$~\%\ of a chance of being real statistically).
However, a partial correlation analysis rejects this trend by giving a value under
the confidence threshold ($\rho_{ML,\Delta\varv} = 0.36$) when keeping
$\Delta\varv$ constant. Second, a weak
correlation may exist between thermal line velocity widths and the luminosity of
the sources ($\rho_{\Delta\varv,L} = 0.49$) but again partial correlation
factors reduce this value to $0.34$, by keeping the mass constant. This result has a 16.6~\%\ probability of being obtained by chance, which is too
high to claim that a link exists. A third correlation appears between luminosity
and mid-IR emission, with $\rho_{L,L_{12}} = 0.80$. This trend is clearly
confirmed
by partial correlation factors, indicating that
we can be confident in this result to a level of more than 99.9\% (see
Fig.~\ref{fig:corr-all}).

\subsection{Methanol correlations}

We show in Sect.~\ref{sec:methmas} that two components can be extracted
from the methanol line emission: a thermal ($A_\mathrm{t}$) and a maser one
($A_\mathrm{m}$). As these two components reflect two different characteristics
of the sources, we try to find correlations between them and the intrinsic
variables (mass, luminosity, etc.). We also test a possible correlation with
the ratio $X_\mathrm{m}$ and the abundance X$_\mathrm{mol}$. The overall correlation
matrix, summarizing the results, is
given in Table~\ref{tab:corr-matrix}.

\begin{table}[t!]
  \begin{center}
  \caption{Correlation matrix for the sample and the CH$_3$OH emission analysis.}
  \label{tab:corr-matrix}
\begin{tabular}{c|cccccccc}
	& M	&  L 	& $\Delta\varv$ & $L_{12}$ 	& $A_\mathrm{m}$ &
$A_\mathrm{t}$ & $X_\mathrm{m}$ & X$_\mathrm{mol}$ \\
\hline
M		& 1 		& 	&	&	&	&	&  &\\	
L		& 0.50 	& 1	&	&	&	&	& & \\
$\Delta\varv$	&\textit{0.45} & 0.49 & 1	&	&	&	& & \\
$L_{12}$	&\textit{0.23} & \textbf{0.80}& \textit{0.19} & 1 & & & & \\
$A_\mathrm{m}$ & \textit{0.34} & \textit{-0.41} & \textit{0.16} & \textbf{-0.67} & 1 & & & \\
$A_\mathrm{t}$ & \textbf{0.70} & \textit{0.01} & \textit{0.31} & \textit{-0.39} & \textbf{0.76} & 1 & & \\
$X_\mathrm{m}$		& \textit{-0.01} & -0.58 & \textit{0.02} & -0.68 & \textbf{0.88} & \textit{-0.37} & 1 & \\
X$_\mathrm{mol}$ & \textit{0.37} & \textit{0.10} & \textit{-0.05} & \textit{-0.01} & \textit{0.21} & 0.51 & \textit{-0.09} & 1 \\
\hline
\end{tabular}\\
\tablefoot{
Factors under the threshold of confidence are given in italic font, correlations
approved by the partial correlation test are in bold font. As the matrix is
symmetric, we have only filled its lower part.
}
\end{center}
\end{table}

\begin{figure*}
\centering
\begin{tabular}{cc}
\includegraphics[width=7.5cm]{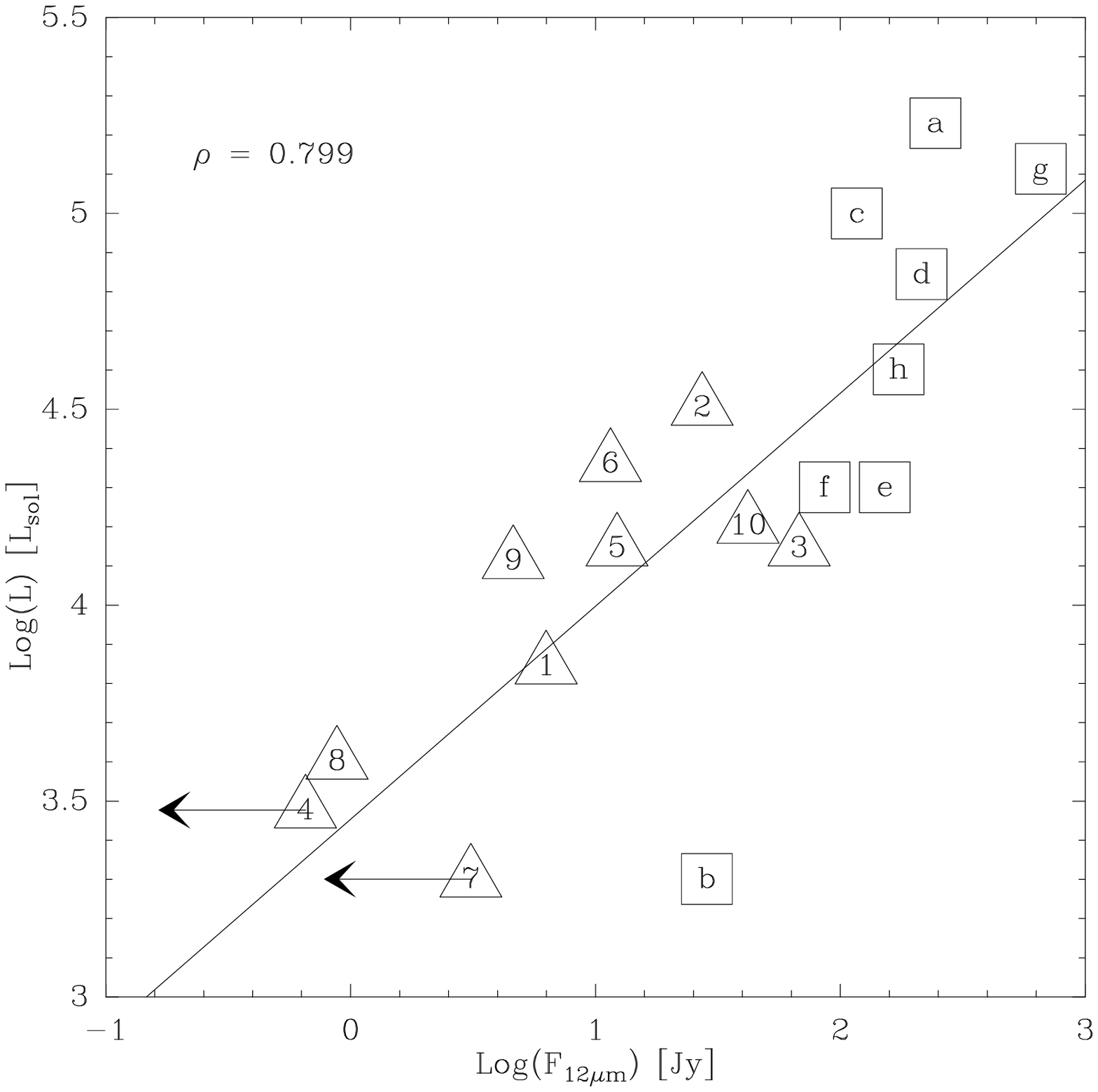}$\mathrm{(a)}$ &
\includegraphics[width=7.5cm]{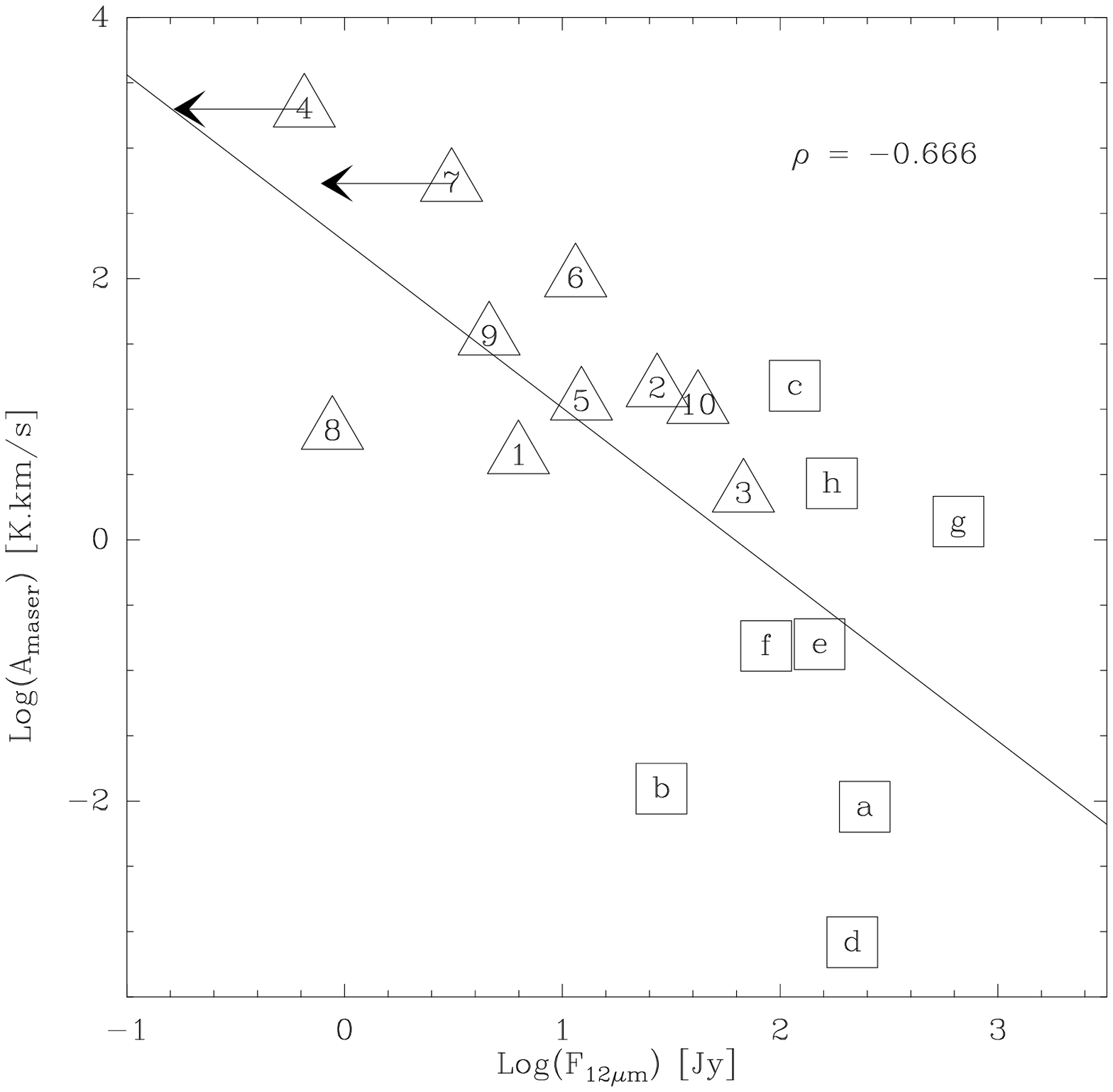}$\mathrm{(b)}$ \\ 
\includegraphics[width=7.5cm]{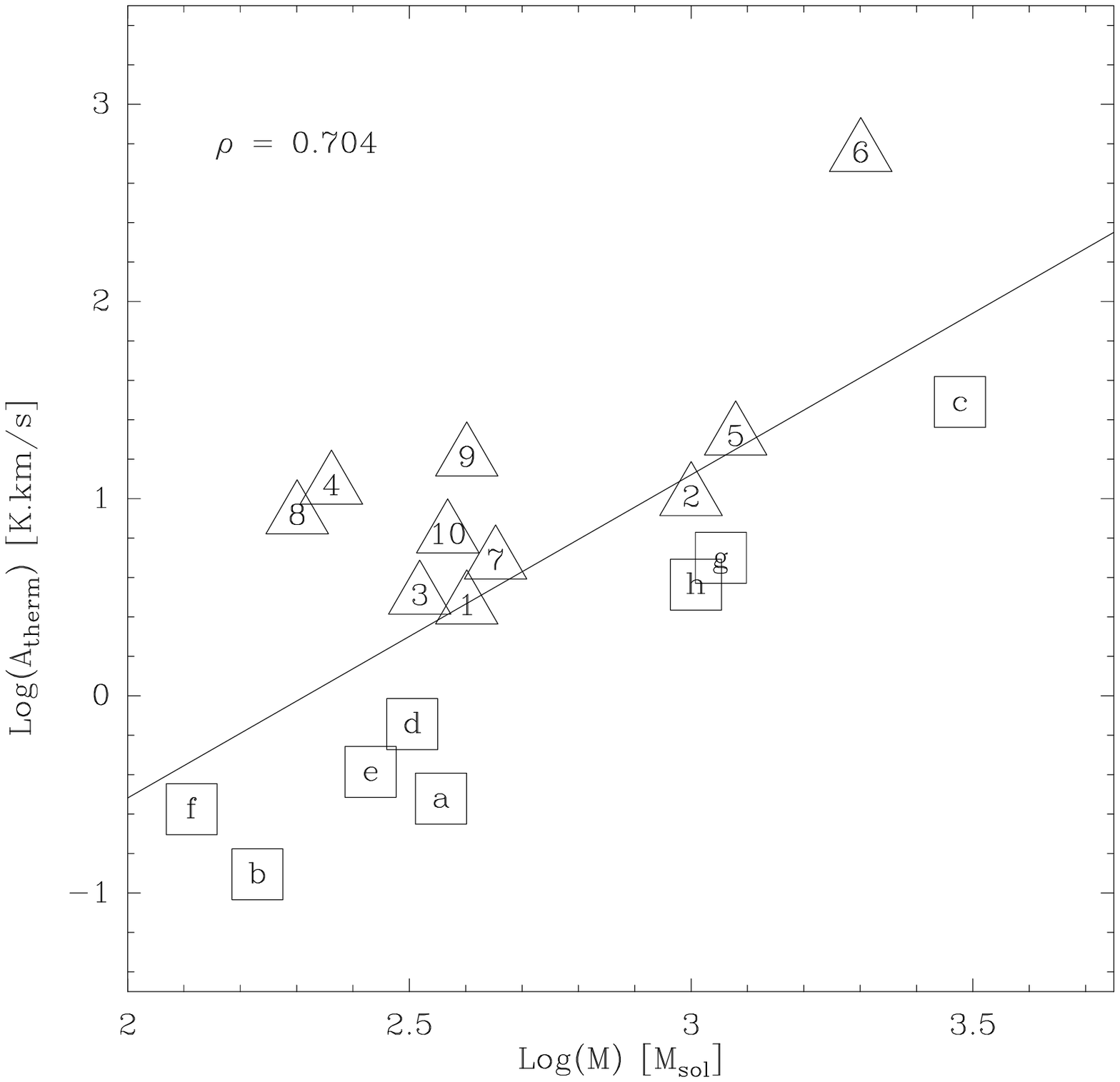}$\mathrm{(c)}$ &
\includegraphics[width=7.5cm]{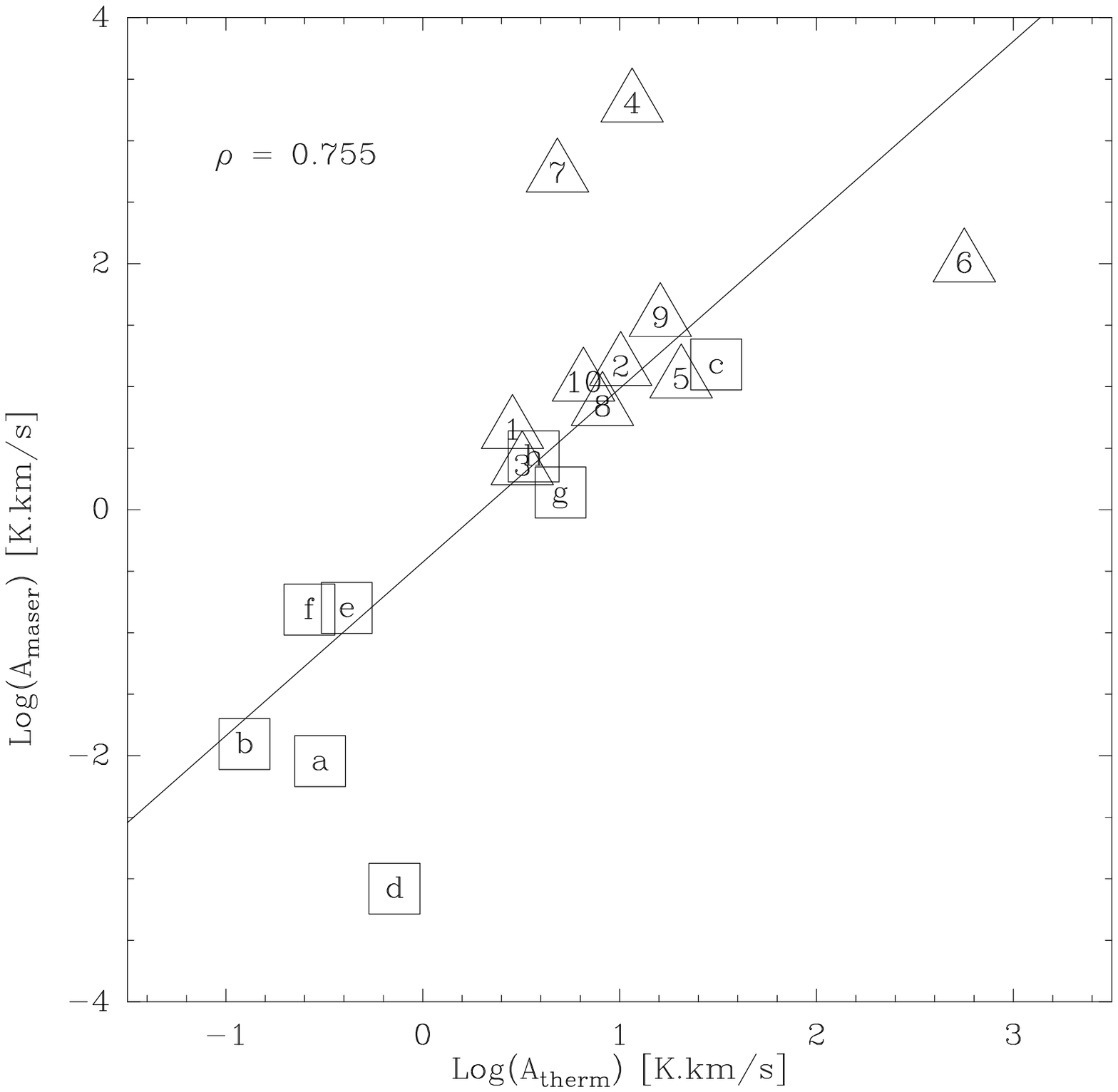}$\mathrm{(d)}$ \\
\includegraphics[width=7.5cm]{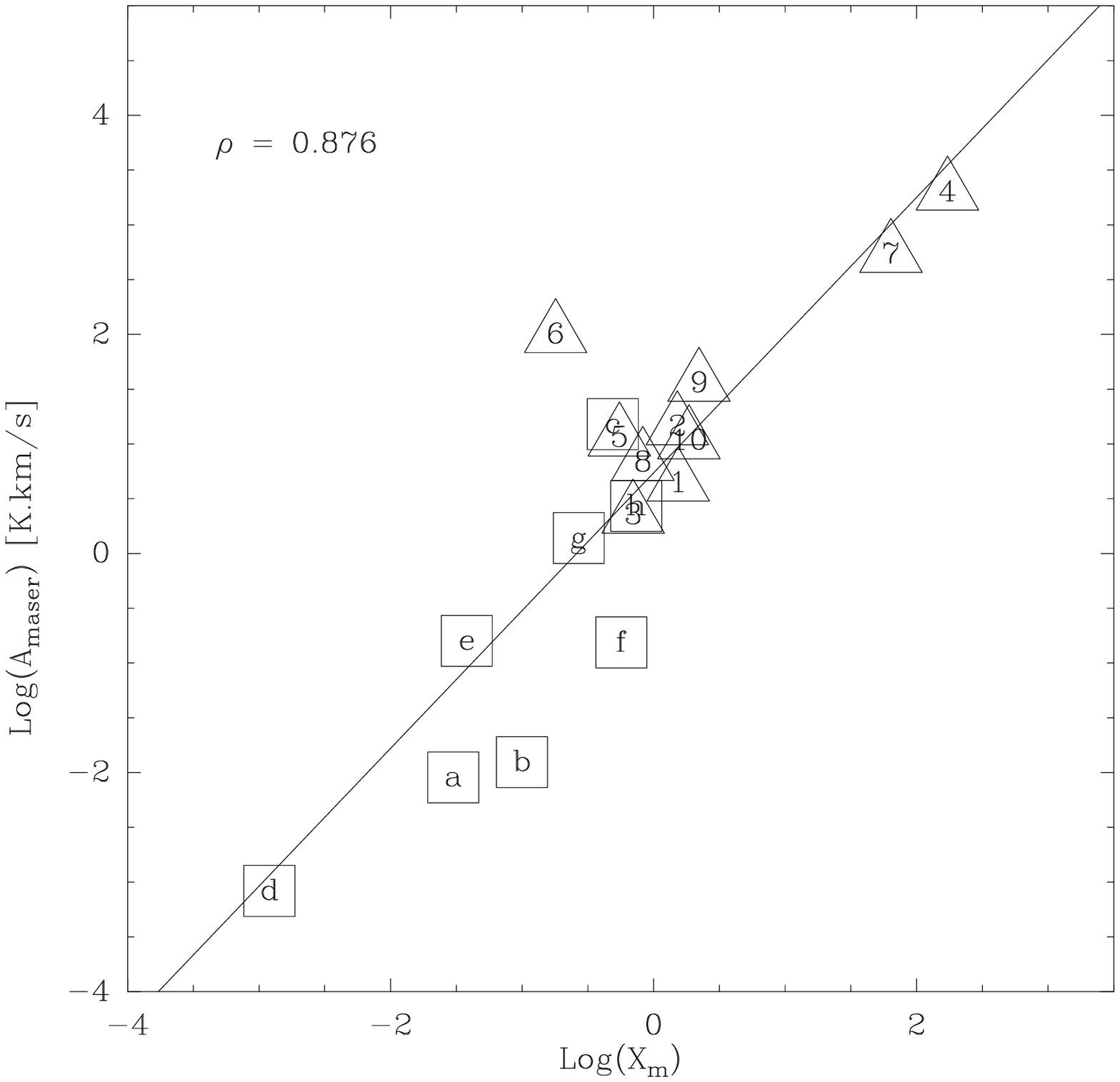}$\mathrm{(e)}$ &
\includegraphics[width=7.5cm]{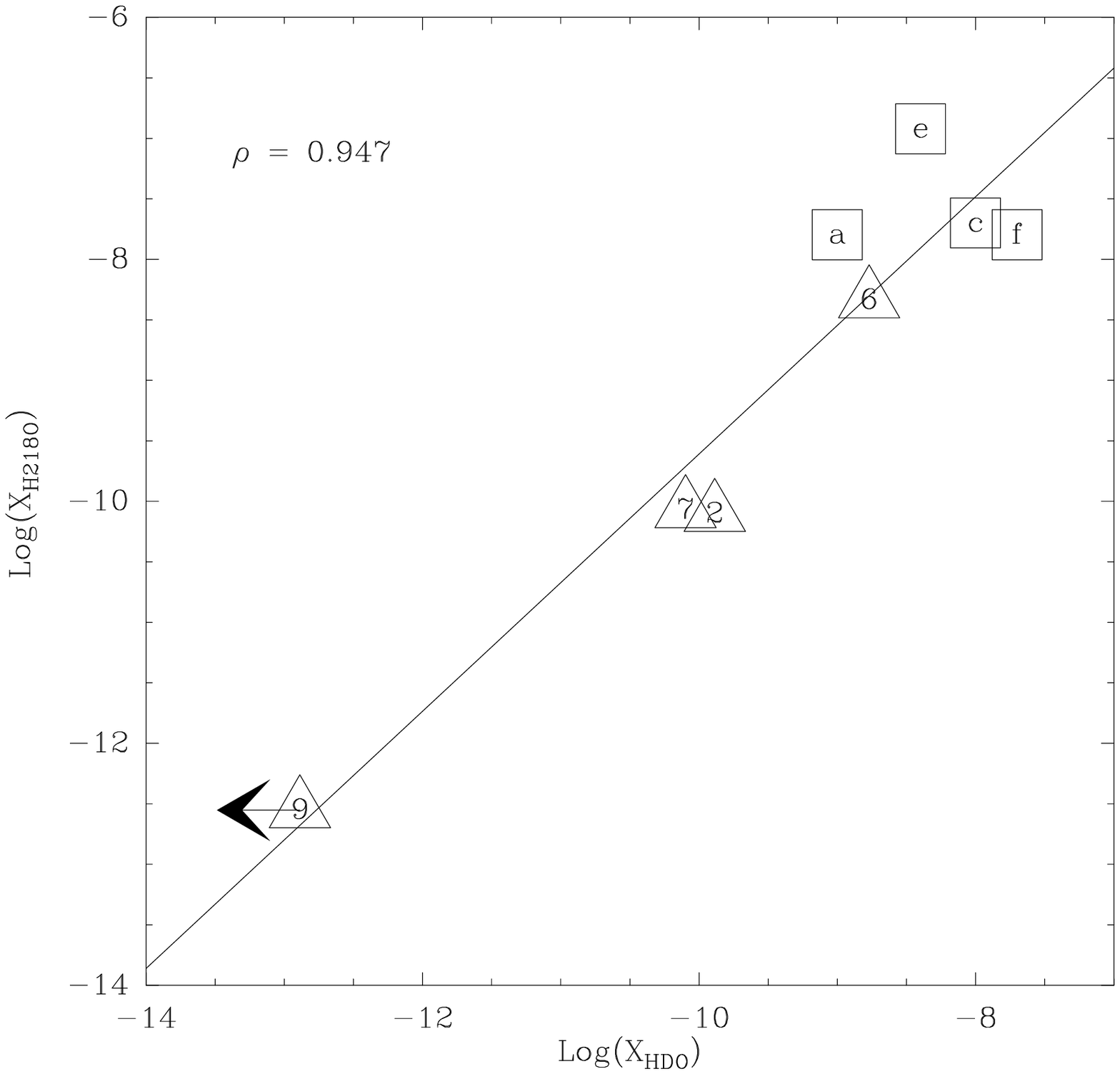}$\mathrm{(f)}$ \\
\end{tabular}
\caption{
Correlation plots between (a) \old{monochromatic luminosities at} 12~\micron\ and source
luminosity, (b) \old{monochromatic luminosities at} 12~\micron\ and maser
component of \CHTOH\ emission line, (c) masses of HMPOs and the thermal
component of
\CHTOH\ emission lines, (d) the thermal and the maser component of \CHTOH\
emission lines, (e) the maser-to-thermal emission ratio and the maser emission, (f) the molecular abundances of HDO and \HDDO. Sources of our 
sample are labelled with numbers (1 to 10), sources
studied by \cite{vandertak2006} are labelled with letters (a to h). Solid line :
best linear fit obtained with a least squares method.}
\label{fig:corr-all}
\end{figure*}

The first correlation relates the mid-IR luminosities to the maser component
($\rho_{A_\mathrm{m},L_{12}} = -0.67$). The correlation factor is negative,
indicating an anti-correlation between these two variables (see
Fig~\ref{fig:corr-all}). This trend is confirmed by the partial correlation
analysis, reducing the correlation factor to -0.65 when the thermal
component is assumed to be constant, and giving approximately a 98.5\%\ of chance
of this correlation being real.
We note that there is no significant correlation of the integrated (thermal + maser) CH$_3$OH line flux in our data.
A partial correlation analysis demonstrates that the weakening of the CH$_3$OH maser flux with increasing 12\,$\mu$m luminosity is not caused by increasing distance.

In addition, we find two correlations associated \old{with} $A_\mathrm{t}$. The first
one indicates that \old{thermal \CHTOH\ emission increases} with the mass of the object 
($\rho_{A_\mathrm{t},M} = 0.70$, see Fig.~\ref{fig:corr-all}). The
partial correlation analysis confirms this trend and, at its most extreme, decreases
the correlation factor to 0.65 when $L$ is assumed to be constant. There is a probability of lower than $0.4$\%\ that this trend is obtained by chance. The thermal emission
is also linked to the maser emission, the correlation factor being equal to
$0.76$ (see Fig~\ref{fig:corr-all}). The study of partial correlation factors
 slightly decreases the initial value to 0.75, giving a high  chance (99.9\%\
statistically) for this result being real. 

Three other links, all related to the ratio $X_\mathrm{m}$ (see
Table~\ref{tab:corr-matrix}), are found but rejected by the partial correlation
analysis, except the correlation with the maser emission which increases (up to 0.96; this result is obvious as $X_\mathrm{m} = A_\mathrm{m}/A_\mathrm{t}$). More precisely, partial correlation factors 
obtained  for other correlations are well below the threshold, with $|\rho| = 0.27$ at the maximum, hence there is more than a 27\%\ chance of obtaining
this result by luck. Thus, this ratio is interesting but a bit less relevant than $A_\mathrm{m}$.

\subsection{Water correlations}

We now consider the water abundances, including numbers obtained in
other MDCs \citep{vandertak2006}. We also search for possible correlations with the
deuteration level D/H. Only the sources where water line emission is detected
are included, hence 8 MDCs. As a consequence, the confidence threshold for a
real correlation is now 0.756, assuming a two-tailed test (or two-$\sigma$ test).

First of all, the analysis of biases in this restricted sample reveals, again, a
link between the mid-IR \old{luminosity} at 12~\micron\ $L_{12}$ and the total luminosity
$L$
(see Table~\ref{tab:corr-matrix2}). Partial correlation factors confirm this
trend with a final correlation factor of 0.79, hence 98\%\ of chance of this
relation being real. This is a strong indication that these two variables are
correlated, even when a smaller sample of sources is chosen.

Another correlation is found, between the mid-IR emission $L_{12}$ and the HDO
abundance ($\rho_\mathrm{L_{12},HDO} = 0.80$, see Table~\ref{tab:corr-matrix2}).
This is not confirmed by the partial correlation analysis, for which we measure a factor beneath
the threshold when the \HDDO\ abundance is assumed to be constant at
$\rho_\mathrm{L_{12},HDO,H_2^{18}O} = 0.58$, hence a 12.9\%\ chance of having
obtained this result with luck, statistically.

The last connection that we find is between the HDO and \HDDO\
($\rho_\mathrm{HDO,H_2^{18}O} = 0.95$, see Table~\ref{tab:corr-matrix2} and
Fig.~\ref{fig:corr-all}). This relation is strong and the partial correlation
method confirms this trend by decreasing, at most, the correlation factor
to 0.90 when the mid-IR flux density $L_{12}$ is assumed to be constant. This
value indicates that our result is real in more than 99.8\%\ of the cases. This
correlation was already noted by eye (Table~\ref{tab:abundances}) and 
confirms that the D/H ratio is of the same order of magnitude in \old{all studied} MDCs. 

More globally, our study illustrates that water abundance (hence detection) does not
depend on a basic characteristic of HMPOs (such as their mass or luminosity) but
needs further investigations in each specific source. This point is discussed in
Sect.~8.2.

\begin{table}[t!]
  \begin{center}
  \caption{Correlation matrix for the water emission analysis.}
  \label{tab:corr-matrix2}
\begin{tabular}{c|ccccccc}
	& M	&  L 	& $\Delta\varv$ & F 	& $\mathrm{HDO}$ &
$\mathrm{H_2^{18}O}$ & D/H \\
\hline
M			& 1 		& 	&	&	&	&
& \\	
L			&\textit{0.28} & 1	&	&	&	&
& \\
$\Delta\varv$		&\textit{0.43} & \textit{0.57} & 1	&	&
&	& \\
F			&\textit{-0.13}& \textbf{0.79} & \textit{0.22}  & 1 & &
& \\
$\mathrm{HDO}$ 		&\textit{0.06} & \textit{0.38} & \textit{-0.22} &
\textit{0.71} & 1 & & \\
$\mathrm{H_2^{18}O}$ 	&\textit{0.00} & \textit{0.46} & \textit{-0.18} &  0.80 
& \textbf{0.95} & 1 & \\
D/H	&\textit{0.102} & \textit{0.10} & \textit{0.02}  & \textit{0.01} &
\textit{0.16} & \textit{-0.11} & 1\\
\hline
\end{tabular}\\
\tablefoot{Factors under the threshold of confidence are given in italic font, correlations approved by the partial correlation test are in bold font. As the matrix is symmetric, we have only filled its low part.}
\end{center}
\end{table}

\section{Discussion}

\subsection{Methanol masers: shocks in embedded objects ?}

Our study reveals a clear link between the mid-IR brightness of the 18 MDCs
(including the sample by \citealt{vandertak2006}) that we have observed and
the methanol maser emission at 84.5~GHz (class I maser). Unlike class II masers,
which originate in static regions near massive protostars, class I masers are
known to occur in extended zones, at some distance from the star formation sites
($\sim$10$^4$~AU), where molecular outflows interact with the surrounding
medium \citep{minier2005}. \cite{cyganowski2009} study 20
massive young stellar objects observed with interferometry to confirm this
behaviour, and claim that methanol class I masers trace the molecular outflows
around them. 

Molecular outflows are indicators of an accretion process, and are known to
decrease in intensity with time as the protostar evolves (low-mass case, see
\citealt{bontemps1996}). Presenting an early stage of
massive star \old{formation,  most of
MDCs harbour powerful outflows}. However, extraction of molecular outflow strength is
biased by the unknown orientation and the \old{possible source multiplicity}, and no clear
\old{drop in strength} can be observed between mIRq- and mIRb-MDCs
\citep{beuther2002c,marseille2008}. Furthermore, the decrease in power is
suspected to occur at a later stage, \textit{i.e.} in MDCs where \old{an} \hii\
region has developed. As a consequence, the variation in the methanol class I
maser emission cannot be linked to a drop in the molecular outflow strength
from mIRq to mIRb sources. Like class II masers, class I maser \old{emission needs} a
very specific environment, with a sufficient \old{spatial extend. Moreover, the detection of multiple
velocity components} implies that the class I maser sites are
\old{localized} at the interface between the molecular outflows and the ambient gas.
We can assume that mIRq sources contain more favourable zones where this kind of
maser emission is more likely to occur.

The origin of the \old{varying} mid-IR {brightnesses} in MDCs remains quite unclear.
On \old{the} one hand, modelling their spectral energy distribution shows that the
mid-IR
emission is critically dependent on the line of sight when molecular outflows
dig a cavity \old{from} which IR emission can escape
\citep{robitaille2006,vandertak2006,marseille2008}. Thus the \old{variations in mid-IR brightness}
 are \old{orientation-dependent} and do not reflect any \old{difference in evolutionary stage}. On the other hand, chemical models and observations of
sulphur-bearing molecules or cold gas tracers such as \ndhp\ seem to indicate that
mid-IR brighter sources are more evolved and \old{warmer, relative} to more quiet ones \citep{reid2007, marseille2008,herpin2009}.

The anti-correlation that we find between the 12~\micron\ luminosity and the
maser emission makes the situation clearer. Indeed, the hypothesis that the
\old{variations in mid-IR brightness are} due to orientation differences cannot explain
this trend, as the maser emission is not affected by the line of sight direction.
Furthermore, the multiple velocity components would be more easily detected if
the molecular outflows, \textit{i.e.} the cavity from where the mid-IR is
emitted, were to follow the line of sight (case of mIRb-MDCs). However, we observe the
opposite trend. A valid interpretation defines mIRq-HMPOs as embedded
objects, in which the envelope mainly interacts with outflows. Here, the
term 'embedded' must be defined to avoid confusion. Our study has found that the
envelope masses of MDCs are not linked to their mid-IR brightness,
hence are not sufficient to determine wether they are embedded or not. Taking
into account the material distribution, the SED modelling of MDCs shows that their
spherically symmetric
distribution \old{is responsible for their} mid-IR emission from MDCs
\citep{robitaille2006,marseille2008}. In this sense, our study shows that
mIRq-MDCs are embedded sources. Bright mid-IR objects may thus have more 
flattened envelopes than weak mid-IR ones.
From an observational point of view, this idea is supported by \old{the} \cite{kurtz2004} study of
the \itwt\ source where maser emission is detected around the source velocity (at $-49.7$~\kms\ 
and $-51.7$~\kms). The conclusion of this study is that this observation can be explained by
accretion shocks near the central object. This could explain also what we observe for \wfot, where 
strong clues of infall toward this source \old{have} been detected \citep{herpin2009}.

\old{Nevertheless, an embedded} phase does not
imply
that they correspond to an earlier stage of evolution. For example, this characteristic can be a
direct consequence of the initial conditions of the MDC formation. However, our
results support \old{a} scenario in which the initial collapsing clump, radially
concentrated, is disrupted by the molecular bipolar outflows during the
accretion phase. \old{Moreover, the mid-IR brightness and evolutionary stage of MDCs are
linked}.

\subsection{The origin of water emission}

Previous results on HDO and \HDDO\ detections in mIRb-MDCs obtained by
\cite{vandertak2006} demonstrated that these objects are not systematically strong
H$_2$O emitters. Our study  has a similar detection rate (just below
$\sim$50~\%) 
and does not show any evidence of an emission enhancement from mIRq to mIRb
sources.
This contradicts the hypothesis that
warmer objects should be stronger emitters. Indeed, a warmer environment should
increase the release of water in the gas phase from the ice on dust grains.
Thus, the origin of the strong line emission of water species in particular
sources cannot be linked to the 'quiet' versus 'bright' scheme. Each source must
be treated individually.

The \old{IRAS18089-1732} MDC is the most luminous object in \old{the mIRq} sample
(3.2\ttp{4}~\lsol), \old{and} is also the source where the turbulence level is the
highest
($\varv_\mathrm{T} = 1.2 - 2.6$~\kms). \old{The discernible methanol class I maser emission in
this object ($X_\mathrm{m} = 1.42$) is indicative of interactions between
the molecular outflows and the envelope}. We have shown that the abundance of
\old{H$_2$O and HDO} is high enough to be detected (see Table~\ref{tab:abundances}). We thus
\old{conlude} that, in \ieit, the water abundance enhancement is caused by the large
amount of energy available from luminosity,  gas turbulence, and molecular
outflows (shocks). In addition, the detection of unexpected dimethyl-ether
emission in the \HDDO\ spectrum suggests that complex chemical components 
are released in the gas phase, including the water ice. We propose that the enhancement originates in a hot molecular core whose presence may also explain why water is
detected in this source.

\old{In \wfot, many molecular rotational
transitions have been detected, covering a wide range of
energy \citep{herpin2009}. The large amount of detected species, even
unexpected complex molecules,} are indicative of an extended hot
molecular core, and \old{make this source} of particular interest. \old{Moreover, this region is suspected
to be in global infall \citep{motte2003,herpin2009}. This very likely explains the very high abundances (see Table 9) in this source: the ice on dust grains must have evaporated, increasing the water abundance to $2$\ttp{-6}}. 

\old{ Water line profiles towards \drto\ exhibit a double velocity component that may be linked to powerful
bipolar outflows,  the  water emission being produced by shocks
between molecular outflows and the massive envelope.} This is confirmed by \old{the strong methanol class I maser emission 
($X_\mathrm{m} = $ 1.58), and wherein three components are identified, but the velocity components observed in H$_2$O and HDO profiles differ} from the
velocity components of the maser emission (see Table~\ref{tab:comp-drto}). \old{In
addition, this scenario (increase in the amount of water species in the gas phase due to shocks) should apply to \idccd, the most
powerful methanol maser emitter ($X_\mathrm{m} = 13.9$), but no HDO and \HDDO\ emission is detected.}
Hence, we conclude that shocks in \drto\ are not the origin of the abundance
enhancement of water. \drto\ is one of three sources where
unexpected detections of dimethyl-ether have been observed, showing that a hot
molecular core takes place \old{there.} This is the best argument to
explain the water abundance in this source. 
 
\old{On the other hand, HDO is detected in the source \nsfs, where there is no
indication of a hot core. Furthermore,}
this object is not a strong methanol class I maser emitter ($X_\mathrm{m} = 0.21$),
showing that no strong shocks occur inside it. The origin of the water detection
is then more difficult to explain. In addition, our modelling of \nsfs\ shows
that the \old{water abundance is not particularly} high compared to upper
limits derived for \old{other sources (see Table~\ref{tab:abundances}), even if this MDC has the
highest temperature of the sample}, well above the others ($\left<T\right> =
38.7$~K). \old{This high temperature} is caused by the small size of this source (0.05~pc) combined with
its
high luminosity (1.3\ttp{4}~\lsol). \old{Therefore,} we suppose that the emission of water
species in \nsfs, while not as abundant as those in hot
cores, is stronger due to this high temperature. \old{However,
the derived abundances are} consistent with
an abundance jump of 1\ttp{4} in the hottest part of \nsfs. Indeed, we 
obtained low but almost equivalent values when considering either low or high energy
level transitions ($\sim 1$\ttp{-13}). \old{Another possibility would be that \nsfs\ harbors a brand new hot core}.
\old{In that case,} primal ices forming on dust grains in the atomic phase of diffuse
clouds 
are strongly deuterated by the formation mechanism
\citep{lipshtat2004,cazaux2008}. 
This could explain why HDO is detected while \HDDO\ is not, these primal ices
being first
desorbed when a hot core starts to develop.

Our study of a large sample of MDCs permits us to conclude that the water
detection in these objects is not linked to their mid-IR brightness,
\textit{i.e.} a different evolutionary stage or a different distribution of
materials.
In addition, it shows that water detection \old{is associated with a} hot molecular
core. This seems reasonable as in the gas phase hot cores are known to release complex molecules that are detected, while ice on dust grains evaporates. In rare cases, such as \nsfs, water line emission may be
detected due to a globally high temperature in the source. 

\subsection{mIRq vs mIRb: review of common and different properties}

Several clues show that mIRq- and mIRb-MDCs have similar properties. First,
previous work on large samples of MDCs indicate that the velocity widths of
emission lines are dominated by \old{gas turbulence} ($\varv_\mathrm{T} =
0.8 - 2.9$~\kms), whose corresponding width \old{is much} higher than the thermal width
\citep[][]{beuther2002a,motte2007}. Our observations confirm this result, and that this high
turbulence may be the consequence of the initial cloud collapse, as suspected
in large-scale filaments (e.g. DR21 in Cygnus X, see \citealt{motte2007}, or the
Ophiuchus main cloud, see \citealt{andre2007}). In this context, the turbulence
is supposed to dissipate by means of ambipolar diffusion while the MDC evolves,
leading to a quasi-static collapse \citep[see][for a review]{wardthompson2007}.
Our study shows that the turbulence does not vary between mIRq- and
mIRb-MDCs, indicating that their evolutionary stages are similar. We also note
that the range of masses in our sample (200$-$2000~\msol) has no impact on the
mid-IR brightness of the sources. To first order, the most massive sources
should have a higher opacity at mid-IR wavelengths. Our study contradicts this
simple view. The emission process of mid-IR is complex in MDCs, and must include
other parameters. Previous work on the SED modelling of MDCs show that the
density distribution plays a major role in this issue
\citep{robitaille2006,marseille2008}. 

Our work has revealed that the water abundance in MDCs is independent of their
mid-IR brightness. The detection ratio in mIRq and mIRb sources is the same
(slightly below 50\%). Water abundances derived are similar, with a wide range
of values in the external cold parts of the objects (between $\sim$1\ttp{-13}
and $\sim$1\ttp{-9}). 
From our study, we conclude that most objects for which water species are detected
contain a more or less extended hot molecular core.
Finally, \old{we have found that mIRq and mIRb sources share}
the same deuteration ratio D/H 
$\sim$(5--40)\ttp{-4}.
As already reported in
\cite{vandertak2006}, this value is \old{one} hundred times higher than the interstellar
ratio, and confirms that the molecular emission observed is coming from
evaporated ices.
 
These results illustrate the difficulty in differentiating between the mIRq
and mIRb classes. 
The origin of this distinction is linked to the class 0 and class I
definitions for low-mass protostars \citep{andre1993}, where the mid-IR
luminosity is compared to the overall luminosity of the sources.
A ratio $L_\mathrm{\lambda < 25\mu m}/L_\mathrm{bol} < 0.3$\% indicates a class
0 object, and a class I object in the opposite case. This definition was
extended to MDCs where a similar behaviour at mid-IR wavelengths was observed,
and 'quiet' sources were suspected to represent an earlier stage of the massive
star formation evolution. 
However, it appears more and more clearly that 
directly applying definitions from low-mass protostars to high-mass star-forming regions is unwise and may not be adapted to correctly transcribe
different evolutionary stages.
This is caused by the multiplicity of
MDCs, that are now identified as massive proto-clusters, but also to the
presence of
powerful outflows, hot molecular cores and suspected infall
\citep{motte2003,vandertak2006,leurini2007a,wardthompson2007,marseille2008,
herpin2009}. Furthermore, we show that the \old{observed variation in mid-IR brightness
seems} to be associated with \old{the} matter distribution around the protocluster,
being more or
less spherically symmetric. 

\old{Interpreting} mIRq sources as more embedded
objects supports the idea that they represent an earlier stage of
the massive star formation. Indeed, with the hypothesis that MDCs start in a
deeply embedded
phase where matter is then flattened by powerful outflows, as the methanol
class I maser 
trend seems to illustrate, the evolutionary sequence from mIRq- to mIRb-MDCs is still
valid.

\section{Conclusions}

We summarize the main conclusions of our study:
\begin{enumerate}
\item Using a correlation method, we have derived a link between
the methanol class I maser emission and the mid-IR luminosity of the MDCs
that we observed. We propose that this link can be explained by the mIRq sources being more deeply embedded, \textit{i.e.} having a more spherically
symmetric distribution
of their envelope material, creating a favourable environment
for the creation of these masers.
 \item The \old{detection of line emission from H$_2$O and HDO is linked to a hot molecular core, independent of the mid-IR brightness.} 
 \item The study of the overall sample has demonstrated that MDCs have many common
properties: their size (0.1~pc), mass ($\sim 200-2000$~\msol), luminosity
($\sim 1$\ttp{4}~\lsol), turbulent velocity of the gas ($\varv_\mathrm{T} =
1-3$~\kms), incidence of hot cores ($\lesssim 50$\%), and deuteration level ($D/H
\sim 5-40$\ttp{-4}). This implies the idea that these objects can be treated as
a single class, and that they constitute a cluster of protostars, some of them
being massive.
 \item The classification of MDCs is complex due to their multiplicity and the
multiple physical processes (independent of their evolutionary stage) that
occur inside, which have a strong impact on observations, \textit{i.e.} hot molecular core
presence, and powerful outflows, suspected collapse. We finally suggest that the
mid-IR 'quiet' and 'bright' classification needs more investigation but is still
relevant when we assume that mIRq sources represent an earlier stage
of massive star formation than mIRb ones.
\end{enumerate}

In the near future, further investigation of the mIRq versus mIRb-MDCs
evolutionary stage problem will be possible, particularly by means of direct water
\old{line} detections with the Herschel-HIFI instrument and the high-angular
resolution of present (IRAM, SMA, CARMA) and future (ALMA) interferometers.
These may provide us with
the final arguments permitting us to find a coherent classification, by age, of the
MDCs, hence allowing us to describe a complete evolution scenario leading to the
formation of massive stars.

\begin{acknowledgements}
We thank the editor, Malcolm Walmsley, for many comments that improved the quality of the manuscript.
\end{acknowledgements}


\bibliography{biblio}
\bibliographystyle{aa}

\appendix

\section{Elements of Correlation theory}
\label{app:corr}

For more detail and a global review of correlation factor usage, we
refer to \cite{rodgers1988}.

\subsection{Correlation factor definition}

\begin{figure}
\centering
\resizebox{0.5\hsize}{!}{\includegraphics{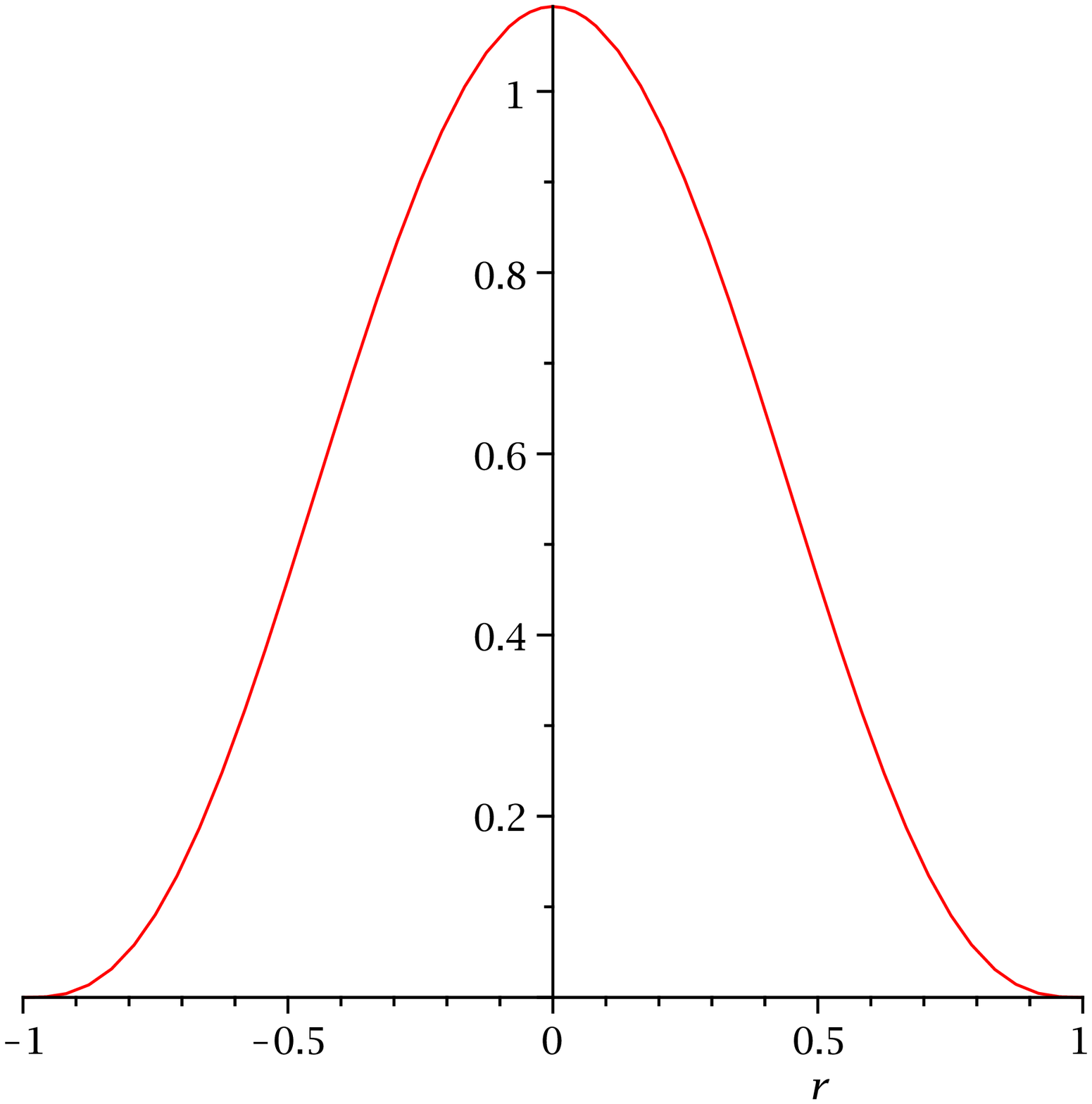}}${(a)}$ \\
\resizebox{0.5\hsize}{!}{\includegraphics{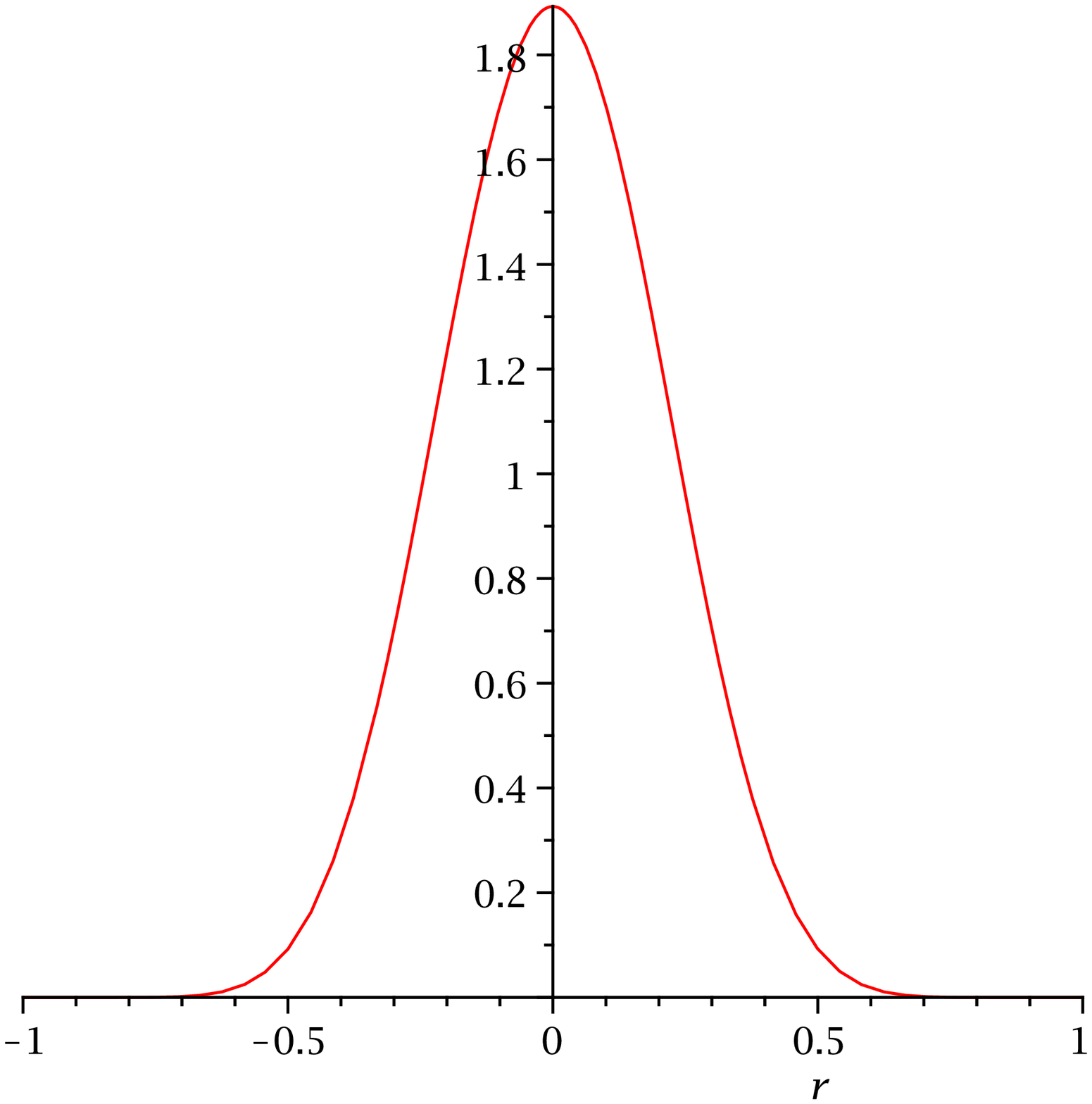}}${(b)}$ \\
\resizebox{0.5\hsize}{!}{\includegraphics{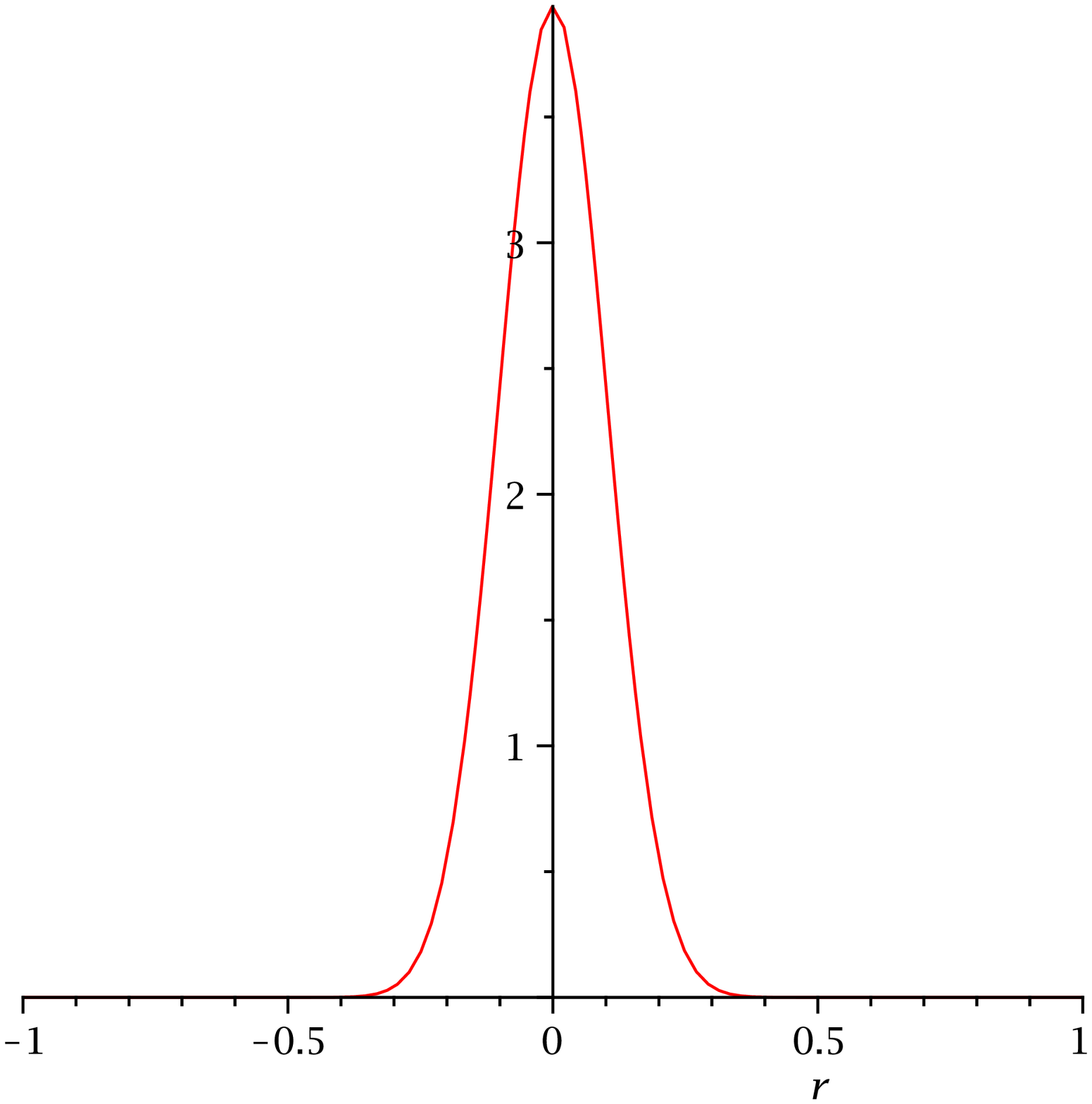}}${(c)}$ \\
\caption{Plots of the $P(\rho)$ with different number of elements on the sample
studied: (a) with $N=10$, (b) with $N=25$, and (c) with $N=100$.}
\label{fig:corr}
\end{figure}

Given two variables $x$ and $y$ with $N$ elements $x_i$ and $y_i$, the
correlation factor $\rho_{x,y}$ between these two variables is measured to be
\begin{eqnarray}
 \rho_{x,y} & = & \frac{\displaystyle\sum_{i=1}^{N}\left(x_i -
\mu_x\right)\left(y_i - \mu_y\right)}{N^2 \sigma_x \sigma_y},
\end{eqnarray}
where
\begin{eqnarray}
 \mu_x & = & \frac{1}{N} \displaystyle\sum_{i=1}^{N} x_i \\
 \sigma_x & = & \frac{1}{N} \displaystyle\sum_{i=1}^{N}\left(x_i -
\mu_x\right)^2.
\end{eqnarray}
A correlation factor has a value between $-1$ and $1$, a positive value
indicating a correlation between the two variables and a negative value
indicating an anti-correlation. If two variables are distributed randomly
following a normal law, the density of probability $P(\rho)$ of finding a
correlation factor between $\rho$ and $\rho+d\rho$ is given by
\begin{equation}
 P(\rho) =
\frac{1}{\sqrt{\pi}}\frac{\Gamma\left(\frac{\nu+1}{2}\right)}{\Gamma\left(\frac{
\nu}{2}\right)}(1-\rho^2)^{(\nu-2)/2},
\end{equation}
where $\Gamma$ is the classic gamma function, and $\nu = N-2$ where $N$ is the
number of elements (see Fig.~\ref{fig:corr} for plots at different $N$). Thus,
the absolute value of the correlation gives, as a function of the number of
elements $N$, a percentage of the distribution $(x_i,y_i)$ obtained by
chance. The higher the absolute value of the correlation factor is, the lower
this percentage. As an example, a correlation factor of $-0.5$ obtained with
$20$ elements has a good chance ($97.5$~\%) of being real. With only $6$ elements,
the same correlation factor has a lesser chance ($68.8$~\%) of being real. 

One can note that the $P(\rho)$ function peaks more and more around $\rho =
0$ while $N$ increases. More precisely, the standard deviation of the $P(\rho)$
distribution is given by
\begin{equation}
 \sigma^2 = <P^2>-<P>^2 = \frac{1}{N-1}.
\end{equation}
It is then common practise to perform a two-tailed (two-sigma) test to reject the
null-hypothesis, \textit{i.e.} that the $(x_i,y_i)$ points does not follow a
distribution of two random variables. This permits us to define a correlation
factor threshold $\rho_\mathrm{t}$ equal to $2\sigma$ to claim wether a correlation
has a good chance of being real or not
\begin{equation}
|\rho| > \rho_\mathrm{t} = \frac{2}{\sqrt{N-1}}. 
\end{equation}
This formula justifies that $N \geq 6$, which is a necessary criterion to obtain
a threshold lower than 1. In the case of our study, we used the two-sigma
test to discriminate between true and false correlation.

\subsection{Partial correlations}

When studying multiple variables at the same time, it is very often
useful to have a clear view of the links between the variables, and first determine
wether they exist or not. Existence of these links can be identified by determining 
the correlation factors of pairs of variables (see section
above).

Unfortunately, the use of correlation factors restricts the study to pairs of
variables, which can lead to the identification of false links, \textit{i.e.} two
variables that seems to be correlated when a third variable is the true
 origin of this link. One can take the example of a sample of pupils of all
ages. A raw study of correlations between their ages, weights, and scientific
levels will certainly make establish a false link: between the weight and the
scientific level. Indeed, we can easily understand that their ages are at the
origin of their growth, physically and intellectually. For a sample
in which this logical deductions would be impossible, how can we distinguish true from false
correlations ?

The use of partial correlations permits us to avoid this issue. Giving a set of
three variables $x$, $y$, and $z$, partial correlation $\rho_{xy,z}$ derives the
correlation factor between two variables (for example $x$ and $y$), assuming that
the third one $z$ remains constant during the measurements. The
partial correlation factor is then given by
\begin{eqnarray}
 \rho_{xy,z} & = &
\frac{\rho_{xy}-\rho_{xz}\rho_{yz}}{\sqrt{(1-\rho_{xz}^2)(1-\rho_{yz}^2)}}.
\end{eqnarray}
If the partial correlation factor obtained is below the confidence threshold
chosen or has a different sign from the standard correlation factor, then
the link between the two variables is false.

For a large set of variables $\left\{X_i\right\}$, all the links
obtained by standard correlations factors $\rho_{X_i,X_j}$ must be verified with
a systematic checking of all partial correlations associated with it,
\textit{i.e.} all $\rho_{X_i X_j,X_{k\neq\{i,j\}}}$ must be above the threshold
of confidence and have the same sign as $\rho_{X_i,X_j}$.
\end{document}